\documentclass[a4paper]{article}
\usepackage{jheppub}

\usepackage{mathabx}
\usepackage{mathtools}
\usepackage{appendix}
\usepackage{slashed}
\usepackage{multirow}
\usepackage{array}
\usepackage{booktabs}
\usepackage{cellspace}
\usepackage{makecell}
\usepackage{braket}
\usepackage{nicematrix}
\usepackage{comment}
\usepackage{hyperref}
\usepackage{subcaption}
\usepackage{cancel,soul}

\usepackage[normalem]{ulem}
 \usepackage{amsmath,amssymb,epsfig,xcolor,bbold,bm, bbm}
 \usepackage{graphicx,caption,subcaption,slashed,tikz}
 \usepackage{hyperref}
 \usepackage{siunitx}
 \usepackage[T1]{fontenc}

\title{Axion Bounds from Quantum Technology}

\author[]{Martin Bauer,}
\author[]{Sreemanti Chakraborti,}
\author[]{and Guillaume Rostagni}

\preprint{IPPP/24/53}

\affiliation[]{Institute for Particle Physics Phenomenology, Department of Physics,\\Durham University,\\Durham, UK}

\emailAdd{martin.m.bauer@durham.ac.uk}
\emailAdd{sreemanti.chakraborti@durham.ac.uk}
\emailAdd{guillaume.rostagni@durham.ac.uk}

\abstract{
A consistent treatment of the quantum field theory of an axion-like particle (ALP) interacting with Standard Model fields requires to account for renormalisation group running and matching to the low-energy theory. 
Quantum sensor experiments designed to search for very light ALPs are particularly sensitive to these effects because they probe large values of the decay constant for which running effects become important. In addition, while linear axion interactions are set by its pseudoscalar nature, quadratic interactions are indistinguishable from scalar interactions. We show how the Wilson coefficients of linear and quadratic ALP interactions are related, including running effects above and below the QCD scale and provide a comprehensive analysis of the sensitivity of current and future experiments. We identify the reach of different experiments for the case of ALP dark matter and comment on how it could be distinguished from the case where it is not the dark matter.  We present novel search strategies to observe quadratic ALP interactions via fifth force searches, haloscopes, helioscopes and quantum sensors. We emphasize the nonlinear behaviour of the ALP field close to the surface of the earth and point out which experimental results are independent on the local background field value.}

\begin{document}
\maketitle
\flushbottom
\clearpage

\section{Introduction}

Quantum sensors have significantly enhanced their sensitivity over the past decade, creating new opportunities to explore fundamental physics questions. These advancements are particularly valuable in testing the effects of light new physics through more precise quantum sensor experiments~\cite{Safronova:2017xyt, Antypas:2022asj}. A prime example of this light new physics is pseudoscalars, which emerge in theories where an approximate global symmetry is spontaneously broken~\cite{Peccei:1977hh,Peccei:1977ur,Weinberg:1977ma,Wilczek:1977pj, Abbott:1982af, Dine:1982ah}. These are referred to as pseudo-Nambu Goldstone bosons, or more commonly, axion-like particles (ALP). The goal of this paper is to translate already performed and anticipated future precision measurements with quantum sensors into a sensitivity range for ALPs. We take the full effective ALP Lagrangian into account, include higher order effects when relevant and perform a consistent calculation of the axion couplings at different scales accounting for renormalisation effects. We stress the importance of quadratic ALP couplings that are formally subleading, but induce spin-independent effects that are strongly constrained from experiments that have not been considered for pseudoscalar interactions.  \\
Light, weakly interacting axions are dark matter candidates~\cite{Preskill:1982cy,Sikivie:2020zpn, Chadha-Day:2021szb}. They can be stable on timescales of the order of the life of the Universe and instead of a production through thermal processes can be produced via the misalignment mechanism. Signals from an axion dark matter background can be very different from a scenario in which axions exist but do not make up a large fraction of the relic dark matter density. In the case that ALPs are dark matter, their quadratic interactions play an important role as they can lead to variations of fundamental constants such as proton, neutron and electron masses and the fine-structure constant. We will enumerate the experimental techniques that can distinguish between the two scenarios and those that are sensitive to both.\\
As a first step, we derive the axion couplings to nucleons, electrons and photons at low energy scales as a function of axion couplings to SM fields. We consider the effects of renormalisation group running and matching to the chiral Lagrangian. We then derive quadratic ALP couplings to nucleons, electrons and photons, which have a significant impact on spin-independent observables. We provide expressions for the dilatonic charges introduced in~\cite{Damour:2010rp} with the ALP couplings in the chiral Lagrangian, allowing to directly connect observables such as variations of fundamental constants and tests of the equivalence principle with the UV theory. We point out that these quadratic ALP couplings lead to an unphysical parameter space close to the surface of the earth due to nonlinear field values sourced by massive bodies~\cite{Hees:2018fpg}. \\
Utilising these results we compare and discuss the sensitivity of different searches for ALPs. Very light bosons induce fifth forces. In the case of ALPs the pseudoscalar couplings result in forces between polarised targets, whereas quadratic ALP exchange induces spin independent forces that fall like $\sim 1/r^4$ with radius $r$. We compare the reach of different experimental techniques and point out the effect from ALP dark matter in this context. \\
We analyse their sensitivity in terms of ALP searches with helioscopes looking to detect ALPs produced in the sun and cavity haloscopes, looking for dark matter ALPs are on the UV couplings and consider a new effect induced by multiple ALP resonances from quadratic ALP couplings.\\
Atomic clock experiments are some of the most sensitive probes of optical transitions. These transitions probe changes in fundamental constants and can therefore test very light ALP dark matter. We again take the effects of renormalisation group induced couplings into account and compare the sensitivity of optical and microwave transitions between hyperfine levels. \\
Laser interferometers are sensitive to shifts in the length and the refractive index of the beamsplitter as expected from ALP dark matter that leads to varying fundamental constants. Atomic interferometers can measure phase differences induced by oscillating electron masses and fine-structure constant in atomic transition frequencies. In both cases, the ALPs can only be detected via the effects induced by their quadratic interactions. 
Mechanical resonators are sensitive to strain in solid objects that can be resonantly enhanced if a dark matter background field oscillates with a wavelength matching the acoustic mode of the resonator. They can be used to probe axion dark matter via its quadratic interactions.\\
In order to illustrate our results we present exclusion contours and sensitivity projections for an example scenario in which the ALP only interacts with gluons in the UV theory, as is the case for the KSVZ QCD axion~\cite{Kim:1979if, Shifman:1979if}. However, our results allow us to translate these constraints and projections for any combination of ALP couplings, extending previous analyses discussing the quadratic ALP-photon coupling~\cite{Kim:2023pvt,Beadle:2023flm}.  

The remainder of this paper is structured as follows: In Section~\ref{sec:theory} we derive the low energy Lagrangian of ALPs including quadratic couplings and their connection to varying fundamental constants in terms of Donoghue's dilatonic charges. Section~\ref{sec:QS} and Section~\ref{sec:probe-quad} contain the discussion of experimental sensitivities to ALPs from searches for fifth searches, BBN, tests of the equivalence principle, haloscopes and helioscopes, atomic clocks, laser and atomic interferometers and mechanical resonators in the case of ALPs as dark matter candidates and the case in which they do not contribute significantly to the relic dark matter density. Section~\ref{sec:summary} gives a summary of the experimental landscape and Section~\ref{sec:conclusions} contains our concluding remarks.
\\

\section{The Low-energy ALP Lagrangian}\label{sec:theory}

At the UV scale axions interact with quarks, gluons and other SM particles. These couplings need to be renormalised consistently and matched to a Lagrangian appropriate for low energy processes. Running and matching changes the axion couplings and introduce new couplings that are not present in the UV theory.

\subsection{Linear interactions} 
At energy scales below the QCD scale $\Lambda_\text{QCD}$ we can write the relevant ALP couplings to photons, nucleons and electrons in the leading order in the expansion in the ALP decay constant $f^{-1}$ as~\cite{Bauer:2021mvw}
\begin{equation}\label{LlowE}
\begin{aligned}
   {\cal L}_{\rm eff}^{D\le 5}(\mu\lesssim\Lambda_\text{QCD})
   &= \frac12 \left( \partial_\mu a\right)\!\left( \partial^\mu a\right) - \frac{m_{a,0}^2}{2}\,a^2\\
      & +\frac{\partial^\mu a}{2f}\,c_{ee}\,\bar e \,\gamma_\mu\gamma_5\, e %
    +g_{Na}\frac{\partial^\mu a}{2f}\bar N \gamma_\mu \gamma_5 N  %
    + c_{\gamma\gamma}^{\text{eff}}\,\frac{\alpha}{4\pi}\,\frac{a}{f}\,F_{\mu\nu}\,\tilde F^{\mu\nu} \,,
\end{aligned}
\end{equation}
where $N =(p,n)$ is a vector containing the proton and neutron spinors, and the linear ALP coupling to nucleons is different for protons and neutrons,
\begin{align}\label{eq:linalp}
g_{Na}=g_0(c_{uu}+c_{dd}+2c_{GG})\pm g_A\frac{m_\pi^2}{m_\pi^2-m_a^2}\bigg(c_{uu}-{c_{dd}}+2c_{GG}\frac{m_d-m_u}{m_u+m_d}\bigg)\,,
\end{align}
where $g_0 = 0.440(44)$ and $g_A=1.254(16)(30)$~\cite{Liang:2018pis, Bauer:2021mvw}, and the positive sign holds for protons and the negative sign for neutrons. The ALP couplings entering \eqref{eq:linalp} are scale dependent and need to be evaluated at the QCD scale, taking into account the effects of running and matching as described in detail in Appendix~\ref{app:running}. 
We give the definition of the ALP couplings in \eqref{eq:linalp} in the UV Lagrangian for completeness
\begin{align}\label{eq:cuucdd}
{\cal L}_{\rm eff}^{D\le 5}(\mu > \Lambda_\text{QCD})\ni \frac{\partial^\mu a}{2f}\,c_{uu}\,\bar u \,\gamma_\mu\gamma_5\, u + \frac{\partial^\mu a}{2f}\,c_{dd}\,\bar d \,\gamma_\mu\gamma_5\, d +c_{GG}\frac{\alpha_s}{4\pi}\frac{a}{f}G_{\mu\nu}\tilde G^{\mu\nu}\,. %
\end{align} 
The relation between the ALP couplings in the UV and of RG running and matching, all ALP couplings to SM particles enter the coefficients in the low-energy Lagrangian \eqref{LlowE} with different strength. The ALP couplings to gauge bosons are defined such that the scale dependence is absorbed by the gauge couplings, such that $c_{GG}$ is not renormalised. The fermion couplings instead at the low scale are sensitive to running and matching contributions, such that  
one finds for $\mu_0=2$ GeV,
    \begin{align}
        g_{pa}(\mu_0) &= 0.88\,c_{GG}(\Lambda)+0.86\, c_{uu}(\Lambda) - 0.42 \,c_{dd}(\Lambda) - 0.40 c_{tt}(\Lambda)\,,\\ 
        g_{na}(\mu_0)  &= 0.012\, c_{GG}(\Lambda)+0.86\, c_{dd}(\Lambda) - 0.42\, c_{uu}(\Lambda) - 0.39\, c_{tt}(\Lambda)\,,\\
        c_{ee}(\mu_0)  &= c_{ee}(\Lambda)-0.002\, c_{GG}(\Lambda)+0.31\, c_{tt}(\Lambda)\,,\label{eq:running1} \\
        c_{\gamma\gamma}^\text{eff}(\mu_0)&=\, c_{\gamma\gamma}(\Lambda) - 1.92\,c_{GG}(\Lambda)
    \end{align}
where additional contributions from the strange, charm and bottom content in the nucleons as well as electroweak running effects from ALP gauge boson couplings and flavor-specific fermion couplings are neglected here. The discrepancy with the results in \cite{GrillidiCortona:2015jxo} or \cite{Vonk:2021sit} is due to these effects, the large contribution from including the top Yukawa coupling, different input parameters and the different choice of Wilson coefficient.\footnote{Relative to these results our ALP gluon Wilson coefficient is defined as $1/f_a=-2\, c_{GG}/f$.} We assume that all ALP-interactions at the low scale are CP conserving, such that it has no linear, scalar coupling to any Standard Model degrees of freedom.

\subsection{Quadratic ALP Interactions}\label{sec:quadraticinteractions}

\begin{figure}[t]
   \centering
\hspace{-2.1cm}\includegraphics[width=0.85\textwidth]{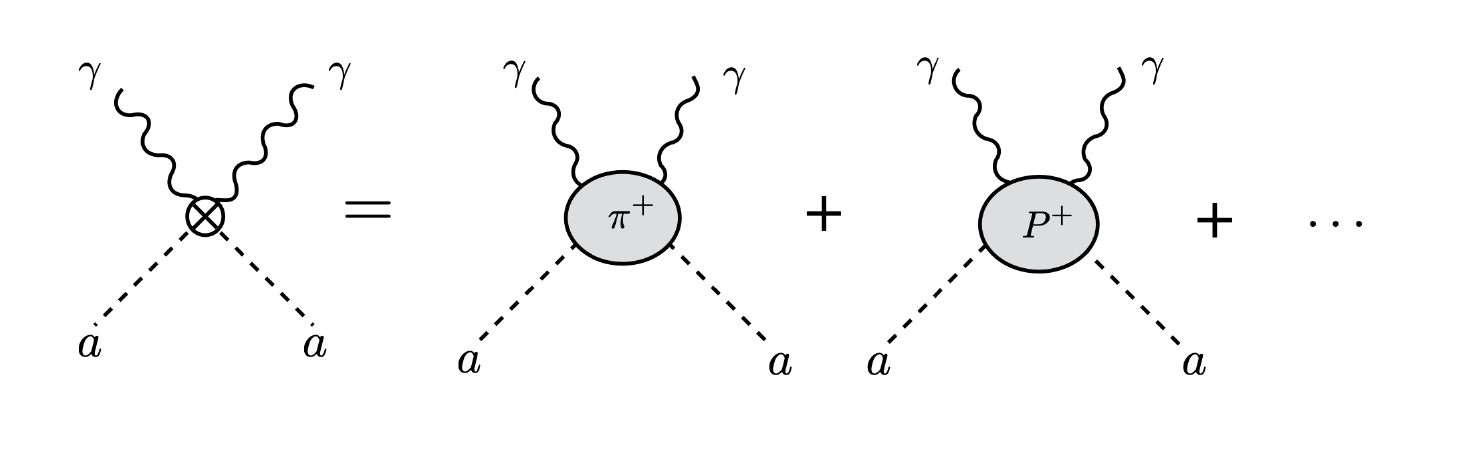}\\[-2em]
\includegraphics[width=\textwidth]{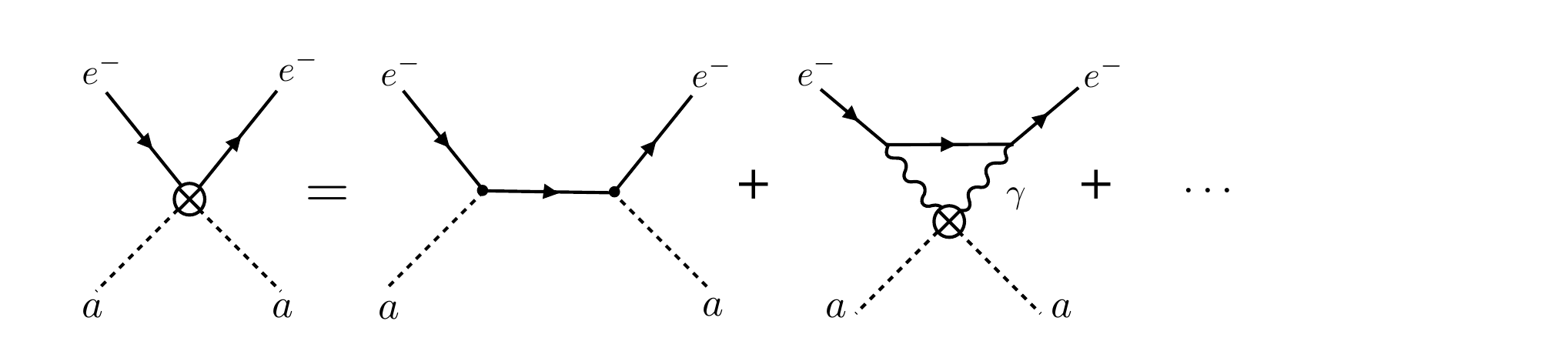}
\caption{\small Diagrams contributing to the quadratic ALP-photon coupling (upper panel) and the quadratic ALP-electron coupling (lower panel).\label{fig:photoncoupling}}
\end{figure}

The ALP Lagrangian in \eqref{LlowE} is linear in the ALP field. At quadratic order in the ALP decay constant ALPs have scalar interactions described by the dimension six operators
\begin{align}\label{a2Lag}
   {\cal L}_{\rm eff}^{D= 6}(\mu\lesssim\Lambda_\text{QCD})
   &=\bar N\left(C_{N}(\mu)\mathbbm{1}+C_{\delta}(\mu)\tau\right)  N \frac{a^2}{f^2} +C_{E}(\mu)\frac{a^2}{f^2} \bar e e  +  C_{\gamma}(\mu) \frac{a^2}{4 f^2} F_{\mu\nu}F^{\mu\nu}\,,
\end{align}
with $\tau=\text{diag} (1,-1)$\,.
These operators induce variations of fundamental constants if the ALP field is light and contributes to the dark matter density. Note that the operators in \eqref{a2Lag} are not invariant under the shift symmetry and as a result, their coefficients are proportional to shift-symmetry breaking terms. In the limit of very small ALP masses $m_a\ll m_q, \Lambda_\text{QCD}$ the main contribution to shift symmetry breaking is the ALP coupling to QCD~\cite{Flambaum:2004tm}. In particular, the pion mass term depends on the quadratic ALP field,
\begin{align}\label{eq:B0term}
\mathcal{L}_{\chi\text{PT}}=\frac{f_\pi^2}{4} \text{tr}[\Sigma m_q(a)^\dagger+ m_q(a)\Sigma^\dagger ] +\ldots\,,
\end{align}
where $\Sigma=\exp\big(i\sqrt2 \Pi/f_\pi\big)$ and the quark mass matrix is ALP-field dependent
\begin{align}
m_q(a)=e^{-i\kappa_q \frac{a}{f}c_{GG}}m_q e^{-i\kappa_q \frac{a}{f}c_{GG}}\!,
\end{align}
where $m_q=\text{diag}(m_u,m_d)$ are the quark masses, and $\kappa_q=\text{diag} (\kappa_u,\kappa_d)$ are unphysical phases subject to the constraint $\kappa_u+\kappa_d=1$. Matching the ALP to the chiral Lagrangian has been performed at the next-to-next-to-leading order level in the case of baryons~\cite{Vonk:2020zfh,Vonk:2021sit} and at next-to leading order for the weak chiral Lagrangian~\cite{Bauer:2020jbp,Bauer:2021wjo, Cornella:2023kjq}. For our purposes, it is sufficient to work in the leading order 2 flavor theory.

The operator \eqref{eq:B0term} induces mass mixing between the ALP and the pion and in the basis where kinetic and mass terms are diagonal one finds upon expanding in $a/f$ that
\begin{align}
m_{\pi,\text{eff}}^2(a)=m_\pi^2\big(1+\delta_\pi(a)\big)\,,\quad
\end{align}
with 
\begin{align}
\delta_\pi(a)&=-\frac{1}{8(1-\tau_a)^2}\frac{a^2}{f^2}\bigg[4c_{GG}^2\Big(1-\tau_a^2-\frac{\Delta_m^2}{\hat m^2}(1-2\tau_a)\Big)\notag\\
&\qquad \qquad \qquad\qquad +4c_{GG}(c_u-c_d)\frac{\Delta_m}{\hat m}\tau_a^2+(c_u-c_d)^2\tau_a^2\bigg]\,\\
&=-\frac{c_{GG}^2}{2}\frac{a^2}{f^2}\bigg(1-\frac{\Delta_m^2}{\hat m^2}\bigg)+\mathcal{O}(\tau_a^2)
\label{Eq:shift_mpi2}
\end{align}
where we introduced $\hat m =(m_u+m_d)/2$ and $\Delta_m=(m_u-m_d)/2$ to make it easier to identify the isospin-breaking terms and $\tau_a=m_a^2/m_\pi^2$. The sign of $\delta_\pi(a)$ plays a crucial role in determining the environmental effects of a massive body influencing the ALP field value. Our result \eqref{Eq:shift_mpi2} agrees with \cite{Kim:2022ype, Kim:2023pvt} and \cite{Beadle:2023flm} in the limit $m_a\to 0$.

For the ALP nucleon coupling the leading order term is generated by the higher order operator
\begin{align}\label{eq:MNcorr}
\mathcal{L}_{\chi\text{PT}}^{(2)}=c_1\text{tr}[\chi_+]\bar N N +\ldots\,,
\end{align}
which results in the nucleon mass 
\begin{align}\label{eq:nuclearmass}
M_N =M_0 - 4c_1 m_\pi^2
\end{align}
where the dimensionful coefficient is $c_1=-1.26(14)\,\text{GeV}^{-1}$~\cite{Alarcon:2012kn} and 
\begin{align}
\chi_+&= 2B_0 \big(\xi^\dagger m_q(a) \xi^\dagger+\xi m_q^\dagger(a) \xi\big)\,,
\end{align}
contains the pion fields $\xi=\sqrt{\Sigma}$. The universal ALP-field dependent correction can be directly calculated from \eqref{eq:MNcorr} or by replacing $m_\pi^2\to m_{\pi,\text{eff}}^2(a)$ in \eqref{eq:nuclearmass}. One can then write the universal quadratic ALP-coupling to nucleons as 
\begin{align}\label{eq:chiraloperator}
c_1\text{tr}[\chi_+]\bar N N= C_{N}\frac{a^2}{f^2} \bar N N + \ldots\,=4c_1 m_\pi^2\, \delta_\pi(a) \bar N N+\ldots
\end{align}
This operator can induce universal, ALP-dependent variations of the nucleon masses
\begin{align}
M_N(a)=M_N\Big(1+\delta_N(a)\Big)\quad \text{with}\quad \delta_N(a)=-4c_1\frac{m_\pi^2}{M_N}\delta_\pi(a)\,.
\label{Eq:shift_mn}
\end{align}
 Besides the universal term, there is also a contribution to the nucleon mass splitting. The relevant term in the chiral Lagrangian reads
\begin{align}
\mathcal{L}_{\chi\text{PT}}^{(2)}=c_5\bar N\left(\chi_+-\frac{1}{2}\text{tr}[\chi_+]\right) N +\ldots\,,
\end{align}
which generates the nucleon mass splitting term
\begin{align}
\Delta M_N=m_N-m_P=4c_5\, m_\pi^2 \,\frac{\Delta_m}{\hat m}+\ldots
\end{align}
The ALP-field dependence can again be obtained by replacing $m_\pi^2\to 
m_{\pi,\text{eff}}^2$, so that
\begin{align}%\label{eq:splitshift}
\mathcal{L}_{\chi\text{PT}}^{(2)}&=C_\delta\bar N\tau N \frac{a^2}{f^2}+\ldots=4c_5\, m_\pi^2\delta_\pi(a) \,\frac{\Delta_m}{\hat m}N\tau N+\ldots
\end{align}
and  the nucleon mass difference in leading order in $\tau_a$ can be written as
\begin{align}\label{eq:splitshift}
\Delta M_N\big(1+\delta_{\Delta M}(a)\big)\quad \text{with}\quad \delta_{\Delta M}(a)=\delta_\pi(a)\,.
\end{align}
 The quadratic ALP couplings to photons are sensitive to charged pion and nucleon loops represented by the second and third diagram on the right-hand side of the upper row in Figure~\ref{fig:photoncoupling}, that can be calculated via threshold corrections to the QED beta function~\cite{Kim:2023pvt} or from the chiral Lagrangian,
\begin{align}\label{eq:Cgamma}
C_\gamma(\mu)&= \frac{\alpha}{24\pi}c_{GG}^2\left(-1+32 c_1\frac{m_\pi^2}{M_N}\right)\left(1-\frac{\Delta_m^2}{{\hat m}^2}\right)
\end{align}
This induces an ALP field-dependent variation of the fine-structure constant
\begin{align}
\alpha^\text{eff}(a)&=\Big(1+\delta_\alpha(a)\Big)\alpha\,\qquad\text{with}\qquad
\delta_\alpha(a)=\frac{1}{12\pi}\left(1-32c_1\frac{m_\pi^2}{M_N}\right) \, \delta_\pi(a) \,.
\label{Eq:shift_alpha}
%\frac{1}{\pi}\left(\frac{2\alpha}{\pi}c_{\gamma\gamma}^2-\frac{2}{3}\frac{m_um_d}{(m_u+m_d)^2}c_{GG}^2+\frac{8}{3}c_1\frac{m_\pi^2}{M_N}c_{GG}^2\right)\frac{a^2}{f^2},
\end{align}
We note that \cite{Beadle:2023flm} find the opposite sign in \eqref{eq:Cgamma}. 
Shift symmetry breaking in quadratic ALP-lepton couplings is a loop effect because the tree-level diagram in the lower row of Figure~\ref{fig:photoncoupling} results in a momentum suppressed contribution to a higher order operator with additional derivatives $(\partial a)^2$.
The leading contribution to the quadratic ALP electron coupling is a result of photon loops~\cite{Kim:2023pvt} 
\begin{align}
 C_E &=-m_e\frac{3\alpha}{4\pi}C_\gamma\ln \frac{m_\pi^2}{m_e^2}
\end{align}
which implies an ALP field-dependent variation of the electron mass 
\begin{align}
m_e^\text{eff}(a)=m_e \big(1+\delta_e(a)\big)\quad\text{with}\quad \delta_e(a)=\frac{3\alpha}{4\pi}C_\gamma\frac{a^2}{f^2}\ln \frac{m_\pi^2}{m_e^2}
\label{Eq:shift_me}
\end{align}
This quantifies the strength with which the ALP field has quadratic interactions with the relevant low energy degrees of freedom.

\subsection{Deriving low energy couplings from the Chiral Lagrangian}

In deriving the quadratic ALP couplings to matter we follow \cite{Damour:2010rp} in which the sensitivity of the different terms in the semi-empirical mass formula to the coefficients of scalar couplings in \eqref{a2Lag} is derived. Here, we calculate these expressions specifically for quadratic ALP interactions and express them in terms of the couplings in the chiral Lagrangian. In \cite{Damour:2010rp} the interaction of scalar fields with a body made from atoms of mass $m_A$ is derived via their dependence on the scalar field, such that
\begin{align}
\alpha_A=\frac{d \ln m_A(\varphi)}{d\varphi}\,,
\end{align}
where we define the dimensionless field $\varphi=a^2/f^2$.\footnote{In~\cite{Damour:2010rp} the coupling is defined via the modified gravitational potential 
\begin{align*}
V(r)=-G\frac{m_Am_B}{r_{AB}}\left(1+\alpha_A\alpha_B\right)\,.
\end{align*}}
The interaction strength can then be written as
\begin{align}
\alpha_A=\alpha_A^\text{RM}+\alpha_A^\text{bind}+\alpha_A^\text{EM}\,.
\end{align}
with atomic number $A$, charge $Z$ and number of neutrons $N$. We use the expression for the rest mass of the nucleons
\begin{align}
    M_N=M_0+\sigma \pm \frac12 \Delta M _N\,,
\end{align}
where one adds the neutron-proton mass difference $\Delta M_N =m_N-m_P$ for the neutron and subtracts it in the case of the proton.
The ALP field-dependent contributions to the rest mass can then be written in terms of the expressions derived in Section~\ref{sec:quadraticinteractions} as
\begin{align}
\alpha_A^\text{RM}&=\frac{1}{m_A} \left[A \frac{\partial \sigma}{\partial \varphi} +\frac12 (N-Z)\frac{\partial \Delta M_N}{\partial\varphi} +Z \frac{\partial  m_e}{\partial \varphi}\right]\,\notag\\
&=\frac{A}{m_A} \Big[ \Big(\sigma+\frac{A-2Z}{2A}\Delta M_N\Big)\delta_\pi  +\frac{Z}{A}  m_e\delta_e\Big]\,\\
&= \Big(0.048 +0.0014\frac{A-2Z}{2A}\Big)\delta_\pi  +5.4\times 10^{-4}\frac{Z}{A}\delta_e\,,
\end{align}
where we have introduced the ALP-field independent couplings $\delta_\pi = \delta_\pi(a=f)$ etc, and used that the sigma term can be related to the nucleon mass via $\sigma=m_\pi^2 \partial M_N/\partial m_\pi^2=-4c_1m_\pi^2\approx 45$ MeV and $\Delta M_N =1.3$ MeV. 

Analogously one can derive the axion-field dependent corrections to the nuclear binding energy and the electromagnetic corrections, which are only sensitive to variations of the pion mass and the fine-structure constant respectively. One can write the results as~\cite{Damour:2010rp}
\begin{align}
\alpha_A^\text{bind}&=\frac{\delta_\pi}{m_A}m_\pi^2\frac{\partial E^\text{bind}}{\partial m_\pi^2}\\
&=\delta_\pi\left[0.045-\frac{0.036}{A^{1/3}}-0.020\frac{(A-2Z)^2}{A^2}-1.42\times 10^{-4}\frac{Z(Z-1)}{A^{4/3}}  \right] \notag \\
\alpha_A^\text{EM}&=\frac{\delta_\alpha}{m_A}\alpha \frac{\partial m_A}{\partial \alpha}=\delta_\alpha\Big(-1.4+8.2\frac{Z}{A}+7.7\frac{Z(Z-1)}{A^{4/3}} \Big)\time 10^{-4}\,.
\end{align}
The interaction strength with an atom with mass number $A$ can then be written in terms of the `dilatonic charges' as 
\begin{align}\label{eq:dilatons}
\alpha_A=\sum_i Q_i\delta_i
\end{align}
with $\delta_i=\delta_\pi, \delta_{\Delta M},\delta_e, $ and $\delta_\alpha$ and the corresponding charges read
\begin{align}\label{eq:dilatoncharges}
Q_{\hat m}&=F_A\left[0.093-\frac{0.036}{A^{1/3}}-0.020\frac{(A-2Z)^2}{A^2}-1.4\times 10^{-4}\frac{Z(Z-1)}{A^{4/3}}\right]\,,\notag\\
Q_{\Delta M}&=F_A\left[0.0017\frac{A-2Z}{A}\right]\,,\notag\\
Q_{e}&=F_A\left[5.5\times 10^{-4}\frac{Z}{A}\right]\,,\notag\\
Q_{\alpha}&=F_A\left[-1.4+8.2\frac{Z}{A}+7.7\frac{Z(Z-1)}{A^{4/3}}\right]\times 10^{-4}\,,
\end{align}
where $Q_{\hat m}$ measures the sum of quark masses corresponding to $\delta_\pi$ and $F_A=\bar m_A/m_A\approx 1$ is the ratio of the mass of the nucleus over $m_A= A M_N$.  
For tests of the equivalence principle we will use the non-universal charges
\begin{align}\label{eq:NUdilatoncharges}
 Q'_{\hat m}&=\frac{0.036}{A^{1/3}}-0.020\frac{(A-2Z)^2}{A^2}-1.4\times 10^{-4}\frac{Z^2}{A^{4/3}}\,,\notag\\
 Q'_{\Delta M}&=0.0017\frac{A-2Z}{A}\,,\notag\\
Q'_{e}&=0\,,\notag\\
Q'_{\alpha}&=7.7\times 10^{-4}\frac{Z^2}{A^{4/3}}\,,
\end{align}
where we also used that $Z/A\approx 1/2$ and $F_A=1$.

\subsection{ALP dark matter}\label{sec:ALPDM}

Axion-like particles are described by spin-0 fields and if they are very light they can have a high occupation number such that the ALP field can be well described by a classical wave~\cite{Hu:2000ke, Hui:2016ltb, Hui:2021tkt}. 
If such light ALPs contribute to dark matter and their relic density is set by the misalignment mechanism this wave would oscillate around the minimum of their potential with an amplitude proportional to the dark matter density $\rho_\text{DM}$,
\begin{align}\label{eq:DMwave}
a(\vec x, t)= \frac{\sqrt{2\rho_\text{DM}}}{m_a}\cos\big(m_a (t +\vec \beta \cdot \vec x)\big)\,.
\end{align}
Here, $|\vec \beta|\approx 10^{-3}$ is the dark matter velocity so that the $\vec x$ dependent term amounts to a random phase. For axions whose mass is related to the decay constant and with negligible interactions with SM fields, one can show that the relic density of ALP dark matter is related to its mass and the decay constant via~\cite{Hui:2016ltb}
\begin{align}\label{eq:reldensity}
\Omega_a\sim 0.1 \left[\frac{10^{-17}\,\text{GeV}^{-1}}{f}\right]^2\left[\frac{m_a}{10^{-22}\, \text{eV}}\right]^{1/2}\,,
\end{align}
and the oscillation begins after nucleosynthesis. ALPs can have explicit shift symmetry breaking terms that are independent of $f$, but we still assume interactions between the ALP and SM fields not to interfere with the misalignment mechanism, so that \eqref{eq:reldensity} is a reasonable estimate. 

One of the most important observables for ultralight dark matter are variations of fundamental constants. They occur when a constant in the Lagrangian becomes field-dependent and the field changes its value. For example, the case of a scalar coupling to the electromagnetic field strength tensor~\cite{Damour:2010rp} 
\begin{align}
\mathcal{L}_{\text{EM} +\phi}=-\frac{1}{4e^2}\left(1-g_\gamma \phi\right)F_{\mu\nu}F^{\mu\nu}=-\frac{1}{4e^2(1-g_\gamma \phi)}F_{\mu\nu}F^{\mu\nu}+\mathcal{O}(g_\gamma^3)\,,
\end{align}
induces a variation of the fine-structure constant
\begin{align}\label{eq:varyalpha}
\alpha^\text{eff}(\phi)=(1+g_\gamma \phi)\alpha\,.
\end{align}
 Using \eqref{eq:varyalpha} and \eqref{eq:DMwave} leads to an oscillating correction to the fine-structure constant that could be observed by any experiment sensitive enough to probe the strength of $g_\gamma$ which is a dimensionful quantity related to the UV scale suppressing the coupling of the scalar $\phi$ to the field-strength tensors. Similarly, scalar couplings between $\phi$ and gluons, quarks and leptons induce time-dependent variations of the strong coupling constant $\alpha_s$, quark masses and lepton masses, respectively. 
 
 Axions or ALPs interact like pseudoscalars, at low energies their interactions are spin dependent. The linear ALP photon interaction 
\begin{align}
\mathcal{L}_{\text{EM} +a}=-\frac{1}{4e^2}F_{\mu\nu}F^{\mu\nu} +c_{\gamma\gamma}\frac{\alpha}{4\pi}\frac{a}{f}F_{\mu\nu}\tilde F^{\mu\nu} 
\end{align}
does not affect the fine-structure constant. Similarly the axial-vector couplings to fermions in \eqref{LlowE} and \eqref{eq:cuucdd} do not lead to variations in the fermion masses. Instead, ALP couplings can lead to variations of dipole moments of nucleons, atoms and molecules~\cite{Graham:2011qk, Stadnik:2013raa, Oswald:2021vtc}.

In contrast, quadratic interactions of an ALP do induce variations of SM couplings and masses. The quadratic ALP-photon interaction can be written as 
\begin{align}
    \mathcal{L}_\text{EM+a$^2$}=C_\gamma \frac{a^2}{f^2}F_{\mu\nu}F^{\mu\nu}
\end{align}
and therefore 
\begin{align}
\alpha^\text{eff}(a^2)=(1+C_\gamma\, a^2/f^2)\alpha\,. 
\end{align}
Since the ALP couples quadratically this variation manifests not just as a time-dependent oscillation, but also as a constant shift if averaged over a time $T\gg 1/m_a$
\begin{align}
\langle a^2\rangle=\frac{2\rho_\text{DM}}{m_a^2} \langle \cos^2{m_a t}\rangle= \frac{\rho_\text{DM}}{m_a^2 }\,.
\label{Eq:time-avg}
\end{align}
\subsection{Quadratic ALP couplings and the ALP background field}\label{sec:QDM}

Light dark matter with scalar couplings is affected by the presence of massive bodies~\cite{Hees:2019nhn}.   
For ALP dark matter, the equation of motion close to a massive body can be written as
\begin{align}\label{eq:eoms}
\left(\partial_t^2 -\Delta+\bar m_a^2(r)\right)a=J_\text{source}(r)+\mathcal{O}(a^3/f^3)
\end{align}
where $J_\text{source}(r)$ is suppressed by CP violating parameters or proportional to $\gamma_5$, and higher order terms in $a/f$ are neglected. The effective mass term can be written as
\begin{align}
\bar m_a^2(r)=m_a^2+\sum_i \frac{Q_i^\text{source} \delta_i}{f^2}\rho_\text{source}(r)\,.
\end{align}
with the dilatonic charges given in \eqref{eq:dilatoncharges}. The solution at a distance $r$ from a spherical, massive body with radius $R_\text{source}$ from with mass $M_\text{source}$, and assuming $J_\text{source}=0$ can be written as 
\begin{align}\label{eq:DMwave2}
a(r, t)= \frac{\sqrt{2\rho_\text{DM}}}{m_a}\cos\big(m_a t\big)\bigg[1- Z(\delta_i) J_{\pm}\left(\sqrt{3|Z(\delta_i)|}\right)\frac{R_\text{source}}{r}\bigg]\,,
\end{align}
such that the ALP field far from the source is given by \eqref{eq:DMwave} (with $\beta\approx 0$),
and 
\begin{align}\label{eq:DMwave3}
Z(\delta_i)=\frac{1}{4\pi f^2}\frac{M_\text{source}}{R_\text{source}} \sum_iQ_i^\text{source}\delta_i\,,
\end{align}
where 
the function $J_\pm(x)=J_{\text{sgn}(Z(\delta_i))}$ depends on the signs of $Z(\delta_i)$, and 
\begin{align}
J_+(x) =\frac{3}{x^3}(x-\tanh x)\,,\\
J_-(x) =\frac{3}{x^3}(\tan x-x)\,,
\end{align}
which means they are sensitive to the sign of the dilaton charges of the source and the ALP interaction terms \eqref{eq:dilatons}. We make the important observation that this combination is always negative in the case of the earth as the source body and an ALP coupled only to $SU(3)_C$ in the UV\,,
\begin{align}\label{eq:alwaysnegative}
\text{only} \quad c_{GG}\neq 0\quad \Rightarrow \quad  \sum_i Q_i^\text{source}\delta_i <0 \,.
\end{align}
This can be inferred from the form of the pion mass shift $\delta_\pi < 0$ as given in \eqref{Eq:shift_mpi2}, which is strictly negative for any sign of $c_{GG}$. In the presence of ALP-quark couplings, this is not guaranteed, but any such contribution is heavily suppressed by the small ALP mass. As a consequence, it follows from \eqref{Eq:shift_mn}, \eqref{eq:splitshift}, \eqref{Eq:shift_alpha} and \eqref{Eq:shift_me} that for all coefficients  $\delta_i<0$,  $i=\pi, M_N, \Delta M, \alpha, e$. At the same time, the charges in \eqref{eq:dilatoncharges} are positive in the case of an ALP interacting with gluons, for which the sole negative contributions are heavily suppressed since the `dilatonic' charges $ Q_{\hat m}, Q_{\Delta M}\gg Q_\alpha, Q_e$.~\footnote{This remains the case for the  opposite sign of $\delta_\alpha$ found by~\cite{Beadle:2023flm}.}  Note that this is different for a scalar with quadratic couplings or even a more general model in which the sign of \eqref{eq:alwaysnegative} is model-dependent. Note also, that \eqref{eq:DMwave2} depends on the boundary conditions of the ALP field~\cite{Banerjee:2022sqg}. 

The function $J_-(x)$ diverges for values of $x\to \pi/2$. For the ratio of the mass and radius of the earth $M_\oplus/R_\oplus\approx 10^{29}\,\text{GeV}^2$,  and all other Wilson coefficients apart from $c_{GG}$ set to zero it follows that this divergence corresponds to an interaction strength of
\begin{align}\label{eq:cGGconstraint}
%\frac{c_{GG}}{f}\gtrsim \left(
%\frac{3}{4\pi}\frac{M_\oplus}{R_\oplus} |Q_{\hat{m}}|\right)^{-1/2}\approx \frac{1}{10^{14}}\,\text{GeV}^{-1}\\
\frac{c_{GG}}{f}\gtrsim \left(\frac{6}{\pi^3}\frac{m_u m_d}{(m_u+m_d)^2}\frac{M_\oplus}{R_\oplus} |Q_{\hat{m}}|\right)^{-1/2}\!\!\!\approx \frac{1}{10^{15}}\,\text{GeV}^{-1}\,,
\end{align}
As a result, the ALP field \eqref{eq:DMwave2} has non-zero field values close to the surface of the earth with crucial consequences for measurements of varying fundamental constants that are sensitive to the ALP field value squared, as well as for experiments measuring nuclear magnetic resonances, electric dipole oscillations~\cite{Hill:2015kva,Chu:2018avg} and atomic spin precession in the ALP background field~\cite{Graham:2017ivz, Abel:2017rtm, Agrawal:2022wjm, Dror:2022xpi}. In contrast to quadratically interacting scalars however, fundamental constants cannot change sign in this parameter space~\cite{Stadnik:2021qyf,Banerjee:2022sqg}. 

For small ALP interactions $c_{GG}/f<10^{-15}$ GeV$^{-1}$, one can expand $J_-(x)=1+\mathcal{O}(x^2)$, such that one can write
\begin{align}\label{eq:DMwaveapprox}
a(r, t)= \frac{\sqrt{2\rho_\text{DM}}}{m_a}\cos\big(m_a t\big)\bigg[1- Z(\delta_i)\frac{R_\oplus}{r}\bigg]
\end{align}
In the following, we will use this approximation, but emphasize that it only applies for very small ALP interactions. 

\begin{figure}[t]
   \centering
\includegraphics[scale=0.4]{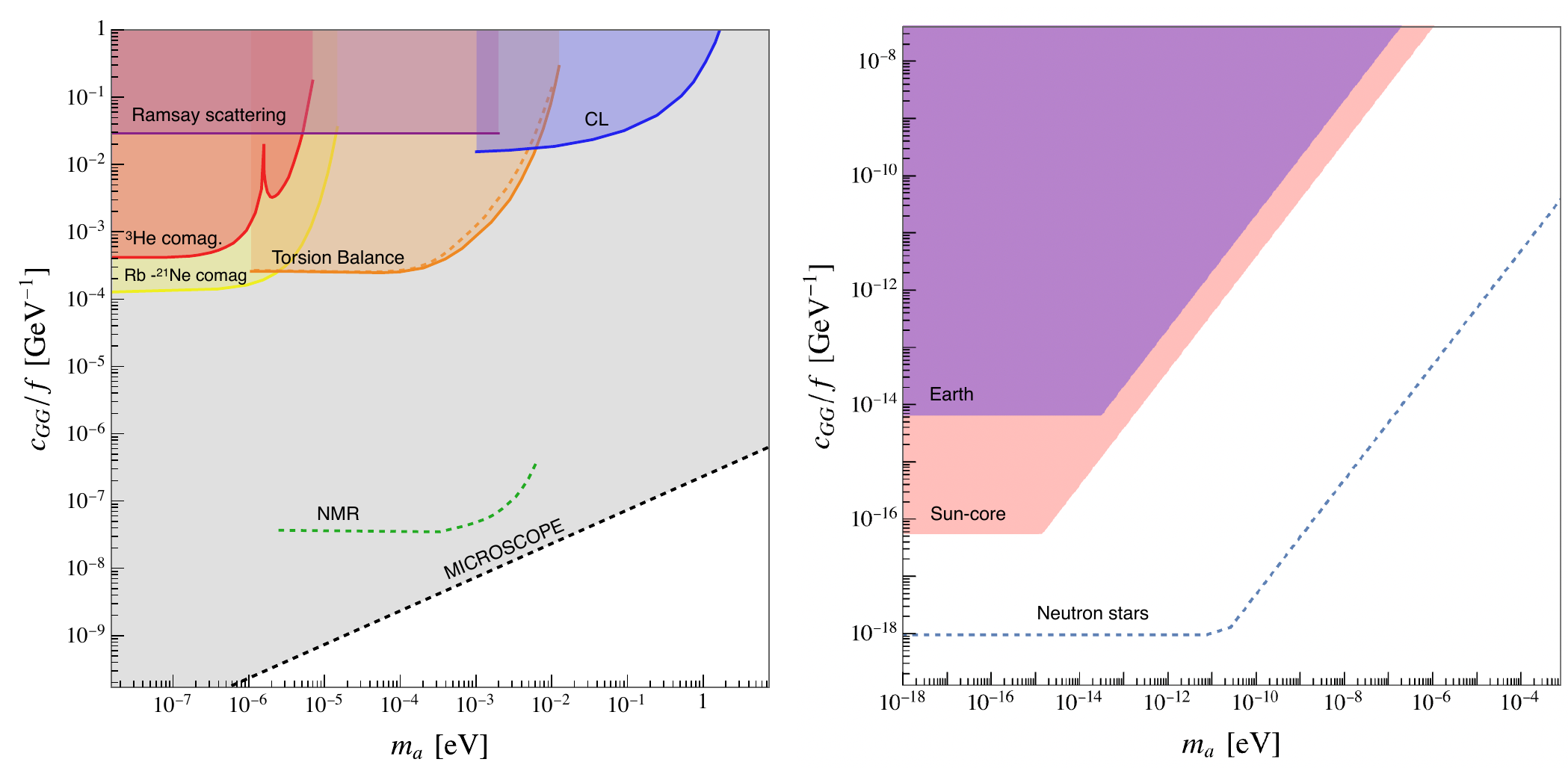}
\caption{\small \label{fig:5force}Bounds from fifth forces on ALP-gluon coupling at the UV scale. Left panel: Constraints from searches for fifth forces induced by ALP exchange in vacuum. Right panel: Parameter space for which the ALP has non-perturbative field values.}
\end{figure}

For large ALP field values instead the full ALP potential needs to be taken into account and higher-order operators regulate the divergence in \eqref{eq:DMwave2}. Instead of \eqref{eq:eoms}, the full equations of motion then reads
\begin{align}
(\partial_t^2-\Delta+m_a^2)\,a&=-\sin\left(\frac{a}{f}\right)\sum_i\,\frac{Q_i^\text{source}\delta_i}{f}\rho_\text{source}(r)\\
&=\left(-\frac{a}{f}+\frac{a^3}{6f^3}-\ldots\right)\sum_i\,\frac{Q_i^\text{source}\delta_i}{f}\rho_\text{source}(r)\,,
\end{align}
and the periodicity of the ALP potential implies a maximal field value of $a\sim \pi f$. For the case of a generic ALP field that is not a dark matter candidate the numerical solution for the full, periodic ALP potential has first been discussed in the context of new constraints from the resulting long-range forces between neutron star inspirals~\cite{Hook:2017psm} and a detailed analysis of the axion phase diagram in systems with finite baryonic density has been performed   in~\cite{Balkin:2020dsr}. A crucial difference between these results and the situation considered here is the boundary condition for the ALP field far from the source. If the axion is not dark matter it vanishes for $r\to \infty$, whereas for ALP dark matter the boundary condition is given by~\eqref{eq:DMwave}. One consequence of this boundary condition is that the dependence of \eqref{eq:DMwaveapprox} on the ALP mass is fixed and the constraint \eqref{eq:cGGconstraint} is ALP mass independent. We nevertheless expect the solution to the equation of motions of the full potential to follow the general argument that the vacuum solution \eqref{eq:DMwave} becomes unstable once the axion mass in the medium is tachyonic ~\cite{Hook:2017psm} 
\begin{align}
m_a^2f^2+\frac{\sigma}{M_N}\rho_N\,\delta_\pi< 0 \,,
\end{align}
where $\rho_N$ is the nucleon density and the corresponding gain in potential energy outweighs the gradient energy$\sim f^2/r^2$ required to perturb the vacuum solution
\begin{align}
\frac{1}{r} \lesssim \Big| m_a^2+\frac{\sigma}{M_N}\rho_N \,\frac{\delta_\pi}{f^2} \Big|^{1/2}\,.
\end{align}
Given these conditions, the QCD axion for which the axion mass is not a free parameter can only deviate from the vacuum solution for densities larger than the nucleon saturation density~\cite{Balkin:2023xtr}. If the ALP mass is treated as a free parameter the parameter space in which the ALP field is sourced by a massive body corresponds to ALP masses below the QCD axion mass, as shown in the right panel of Figure~\ref{fig:5force}. An explicit example of such a model has been discussed in \cite{DiLuzio:2021pxd, Balkin:2022qer}.

% \clearpage

\section{Axion Bounds from Quantum Technology}\label{sec:QS}

In the following we calculate the different contributions to tests of the equivalence principle and searches for fifth forces from ALP interactions. We further compare haloscope and helioscope searches and emphasize the role of the screening effects on their sensitivity. For the calculations presented here we consistently work in the small coupling limit, but clearly denote the parameter space for which this assumption is not justified. We present our results in a model with all UV Wilson coefficients equal to zero apart from $c_{GG}$.  

\subsection{Fifth forces}

The exchange of very light ALPs can induce a fifth force between macroscopic objects. In contrast to a vector boson or a scalar linear ALP interactions are pseudo-scalar and therefore lead to spin-dependent forces~\cite{Moody:1984ba}. Based on the general linear ALP couplings to nucleons 
\begin{align}
\mathcal{L}=g_sa\bar N N +g_{p} a \bar N i\gamma_5 N\,,
\end{align}
there are three types of potentials for the force between two nucleons~\cite{Daido:2017hsl}
\begin{align}\label{eq:linpotentials}
V_{ss}(r)&=-\frac{g_sg_s}{4\pi r}e^{-m_ar}\approx -\frac{g_sg_s}{4\pi r}\,,\\
V_{sp}(r)&=\frac{g_sg_p}{4\pi M_N}(S_2\cdot \hat r)\left(\frac{m_a}{r}+\frac{1}{r^2}\right)e^{-m_a r} \approx\frac{g_sg_p}{4\pi M_N}\frac{S_2\cdot \hat r}{r^2}\,, \\
V_{pp}(r)&=-\frac{g_pg_p}{4\pi M_N^2}\bigg[S_1\cdot S_2\left(\frac{m_a}{r^2}+\frac{1}{r^3}+\frac{4\pi}{3}\delta^{(3)}(r)\right)\\
&\qquad \qquad -S_1\cdot\hat r S_2 \cdot r
\left(\frac{m_a^2}{r}+\frac{3m_a}{r^2}+\frac{3}{r^3}\right)\bigg]e^{-m_a r}\notag\\
&\approx -\frac{g_pg_p}{4\pi M_N^2 r^3}\bigg[S_1\cdot S_2 -3S_1\cdot\hat r \bigg]\,,\notag
\end{align}
where $S_i=\sigma_i/2$ denotes the spin operator of the nucleon, $M_N$ is the nucleon mass and the approximations hold for vanishing ALP mass. In the absence of CP violation, the axion has purely pseudoscalar couplings to SM fermions and $V_{ss}(r)=V_{sp}(r)=0$. However, CP violation in the weak sector or a non-zero theta angle would induce a scalar ALP coupling to quarks $g_s=\theta_\text{eff}\,  m_q/f$. In axion models, the same parameter also appears as a CP-odd quark mass term that is not suppressed by $m_q/f$. As a result the most stringent bounds on $\theta_{eff}\lesssim 10^{-10}$ arise from electric dipole measurements and dominate over bounds from searches for fifth forces by orders of magnitude~\cite{Mantry:2014zsa, Mantry:2014eya}.  

The potential induced by the pseudoscalar ALP couplings is experimentally challenging because they vanish for averaged spins. In the absence of any CP-violating linear ALP couplings the dominant spin-independent contribution to the potential comes from the exchange of pairs of ALPs between nucleons
\begin{align} \label{eq:potential_spur}
    V_\text{2}(r) &= \frac{3g_{Na, 1}^2 g_{Na,2}^2}{128\pi^3 f^4}\frac{1}{r^5}\bigg[\left( x_a + \frac{x_a^3}{6}  \right) K_1(x_a) + \frac{x_a^2}{2}  K_0(x_a) \bigg] \\
    &+ \frac{1}{64 \pi^3 f^4}   \bigg\{ -C_{N_1} C_{N_2}\frac{1}{r^3} x_a K_1(x_a)\notag\\
    &+ \frac{3}{4} \left[    C_{N_1} g_{N_2}^2\frac{1}{m_{N_2}} +   C_{N_2}g_{N_1}^2\frac{1}{m_{N_1}}  \right]  \frac{1}{r^5}  \bigg[\left( x_a + \frac{x_a^3}{6}  \right) K_1(x_a) + \frac{x_a^2}{2}  K_0(x_a) \bigg]\bigg\}\,\notag\\
    &\approx -\frac{C_{N_1} C_{N_2}}{64\pi^3 f^4}   \frac{1}{r^3} \,, \notag
\end{align}
where the interaction strengths are given by \eqref{eq:linalp} and \eqref{eq:chiraloperator}, respectively, and the last line is the approximation in the limit $m_a\to 0$. In order to evaluate the force acting between macroscopic bodies and accounting for all ALP couplings one can replace $C_{N_i}\to \alpha_i$ with $\alpha$ defined in \eqref{eq:dilatons}. Purely derivative ALP couplings lead to forces scaling as $1/r^5$ or $1/r^7$~\cite{Ferrer:1998ue, Bauer:2022rwf}.
 
Another important effect is the acceleration experienced by test masses close to massive bodies like planets, e.g. tests of violations of the equivalence principle. Forces caused by axion fields in vacuum are captured by the potentials \eqref{eq:linpotentials} and \eqref{eq:potential_spur}. 
The acceleration of a test mass $m_T(a)$ in the ALP background field $a(x)$ follows from the equations of motion from the low energy Lagrangian
\begin{align}
L_T=-m_T(a)\sqrt{-g_{\mu\nu} \frac{dx^\mu}{dt}\frac{dx^\nu}{dt}}\approx -m_T(a)\left(1-\frac{\beta^2}{2}\right)\,,
\end{align}
such that in the realm of validity of \eqref{eq:DMwaveapprox}
\begin{align}
\ddot x&= -\frac{2\alpha_T}{f^2} a(x)\,\left(\dot a(x) \dot x +\nabla a(x)\right) \label{eq:accquad}\\
&\approx \frac{\alpha_T}{\pi f^4}\frac{\rho_\text{DM}}{m_a^2}\sum_i Q_i\delta_i\frac{M_\text{source}}{r^2}  \bigg(1-\frac1r\frac{M_\text{source}\sum_i Q_i\delta_i}{4\pi f^2}\bigg)\cos^2(m_a t)\,,\notag\\
&\approx \frac{\alpha_T}{\pi f^4}\frac{\rho_\text{DM}}{m_a^2}\sum_i Q_i\delta_i\frac{M_\text{source}}{r^2} \cos^2(m_a t)\,,
\end{align}
neglecting contributions suppressed by $\beta$ and the last line neglects higher order terms in $1/f^2$.

In terms of the E\"otv\"os parameter, the relative acceleration between the bodies A and B in the field of the gravitational field of a source is given by
\begin{align}
\eta=2\frac{|\ddot x_A-\ddot x_B|}{|\ddot x_A+\ddot x_B|}= 8\frac{\rho_\text{DM}}{m_a^2}\frac{M_\text{Pl}^2}{f^4}\sum_{i,j} Q_i^\text{source}\delta_i \big(Q_j^A\delta_j  -Q_j^B\delta_j  \big)\,.
\end{align}
Tests of the equivalence principle are particularly sensitive to non-universal ALP couplings. We stress again that parts of this parameter space is excluded due to the nonlinearity of the ALP field close to earth which are shown for the case that the axion isn not dark matter in the right panel of Figure~\ref{fig:5force}. Future tests with muonic atoms could significantly improve the sensitivity of fifth-force searches~\cite{Stadnik:2022gaa}.  \\

In Figure~\ref{fig:5force} we compare constraints from different searches for fifth forces induced by ALP exchange assuming a model with one coefficient, $c_{GG}$, in the UV. The left panel shows the constraints from searches for so-called {\fontfamily{qcr}\selectfont Casimir-less forces} shown in blue~\cite{Klimchitskaya:2015zpa, Bauer:2023czj}. The dominant contribution to this force is induced by the spin-independent potential \eqref{eq:potential_spur}.  {\fontfamily{qcr}\selectfont Torsion balances} probe the same spin-independent force and the constraint from~\cite{Adelberger:2006dh} is shown in orange. Spin-dependent (dipole-dipole) interactions have been probed with  {\fontfamily{qcr}\selectfont comagnetometers} using spin-polarised K and $^3$He atoms~\cite{Vasilakis:2008yn} and $^{87}$Rb~\cite{Almasi:2018cob}. The purple region labeled  {\fontfamily{qcr}\selectfont Ramsay scattering} is excluded from measurements of spin-dependent forces in molecular H$_2$~\cite{Ramsey:1979bzw}. Future measurements of spin-dependent interactions exploiting  {\fontfamily{qcr}\selectfont nuclear magnetic resonances (NMR)} can significantly improve the current sensitivity as shown in green dashed line in Figure~\ref{fig:5force}~\cite{Arvanitaki:2014dfa,Aybas:2021nvn}. Note, that nuclear magnetic resonances are sensitive to CP-violating ALP couplings.
In the case the ALP constitutes dark matter, the strongest constraint from fifth force searches are obtained by searches for violations of the equivalence principle. The  {\fontfamily{qcr}\selectfont MICROSCOPE} collaboration measured the force necessary to keep two test masses in relative equilibrium on a satellite orbiting earth ~\cite{Touboul:2017grn, Touboul:2022yrw}. The corresponding bound is shown in gray in Figure~\ref{fig:5force}, but we emphasize that this parameter space is affected by the non-perturbative  axion field value discussed in Section~\ref{sec:QDM}. In the right panel of Figure~\ref{fig:5force} shows these constraints from non-perturbative ALP field values following \cite{Hook:2017psm, Balkin:2020dsr}.

\subsection{Haloscopes and Helioscopes}
Searches for axions and ALPs with haloscopes and helioscopes are sensitive to the interactions of ALPs with a magnetic or electric external field. We present limits and projections for existing and future experiments in terms of the ALP coupling to gluons in the UV by taking all renormalisation effects into account and for a wide range of ALP masses. Quadratic ALP interactions can induce a novel signal for which we provide projections and discuss future opportunities. 

\subsubsection{Haloscope Searches for ALPs}
The ALP-photon coupling can be probed in resonant cavities. Haloscopes are microwave cavities tuned to detect the resonant conversion of dark matter ALPs into photons in the presence of a strong static magnetic field. ALP conversion inside the cavity takes place through ``Primakoff production" which is primarily induced by the linear ALP-photon interaction as considered in the low-energy Lagrangian in \eqref{LlowE}. The interaction is proportional to 
\begin{align}\label{eq:cav1}
  c_{\gamma\gamma}^{\text{eff}}\,\frac{\alpha}{4\pi}\,\frac{a}{f}\,F_{\mu\nu}\,\tilde F^{\mu\nu}= c_{\gamma\gamma}^{\text{eff}}\,\frac{\alpha}{\pi}\,\frac{a}{f}\,\vec{E}\cdot \vec{B}\,.
\end{align}
The ALPs with a frequency $\nu_a$ convert to photons if $\nu_a$ matches the frequency of a resonant mode of the cavity resonator. The frequency follows the relation $h\nu_a=m_a\,c^2+1/2\, m_a v_a^2$, where $v_a$ is the dark matter velocity dispersion in the galactic halo~\cite{Stern:2015kzo,McAllister:2023ipr}. Photons generated from ALP-photon conversion give rise to excess power generation inside the cavity. The signal power extracted on resonance is given by
\begin{align}
    P_{a\to\gamma} =
    \frac{\alpha^2}{\pi^2}\,\frac{\big(c^\text{eff}_{\gamma\gamma}\big)^2}{f^2}\,\frac{\rho_\text{DM}}{m_a} B_0^2 V C\, {\rm min}(Q_L,Q_a)
    \label{eq:halo1}
\end{align}
where $B_0$ denotes the field strength of the external magnetic field, $C$ is a mode-dependent form factor that quantifies the overlap between the EM fields of the cavity mode and the mode induced due to ALP-photon conversion. $V$ denotes the cavity volume, $\rho_{\rm DM}$ is the local dark matter density (all haloscopes assume ALPs comprise 100\% of the dark matter density). The min function picks the smaller number between the loaded cavity quality factor ($Q_L$) intrinsic to the cavity material and the ALP signal quality factor ($Q_a$). $Q_a$ scales with the velocity dispersion as $Q_a \sim 1/\langle v_a^2 \rangle \sim 10^6$. This implies that for shorter timescales than the ALP coherence time $\tau_a\sim Q_a/m_a$, one can treat it as a monochromatic field as in \eqref{eq:DMwave}.  
The bound on the ALP-photon coupling can be obtained by setting a target value of the signal-to-noise ratio (SNR) which is given by
\begin{align}
   {\rm SNR} = \frac{P_{a\to\gamma}}{k_B T_\text{exp}}\sqrt{\frac{t_\text{exp}}{\Delta\nu_a}}\,,
\end{align}
where the system temperature $T_\text{exp}$ is the system noise temperature, $t_\text{exp}$ is  the total integration time and $\Delta\nu_a$ corresponds to the ALP signal linewidth which scales as $\Delta \nu_a~\approx~m_a\, \langle v_a^2 \rangle$~\cite{ADMX:2019uok}.

\begin{figure}[htb!]
 \centering
 \includegraphics[width=\textwidth]{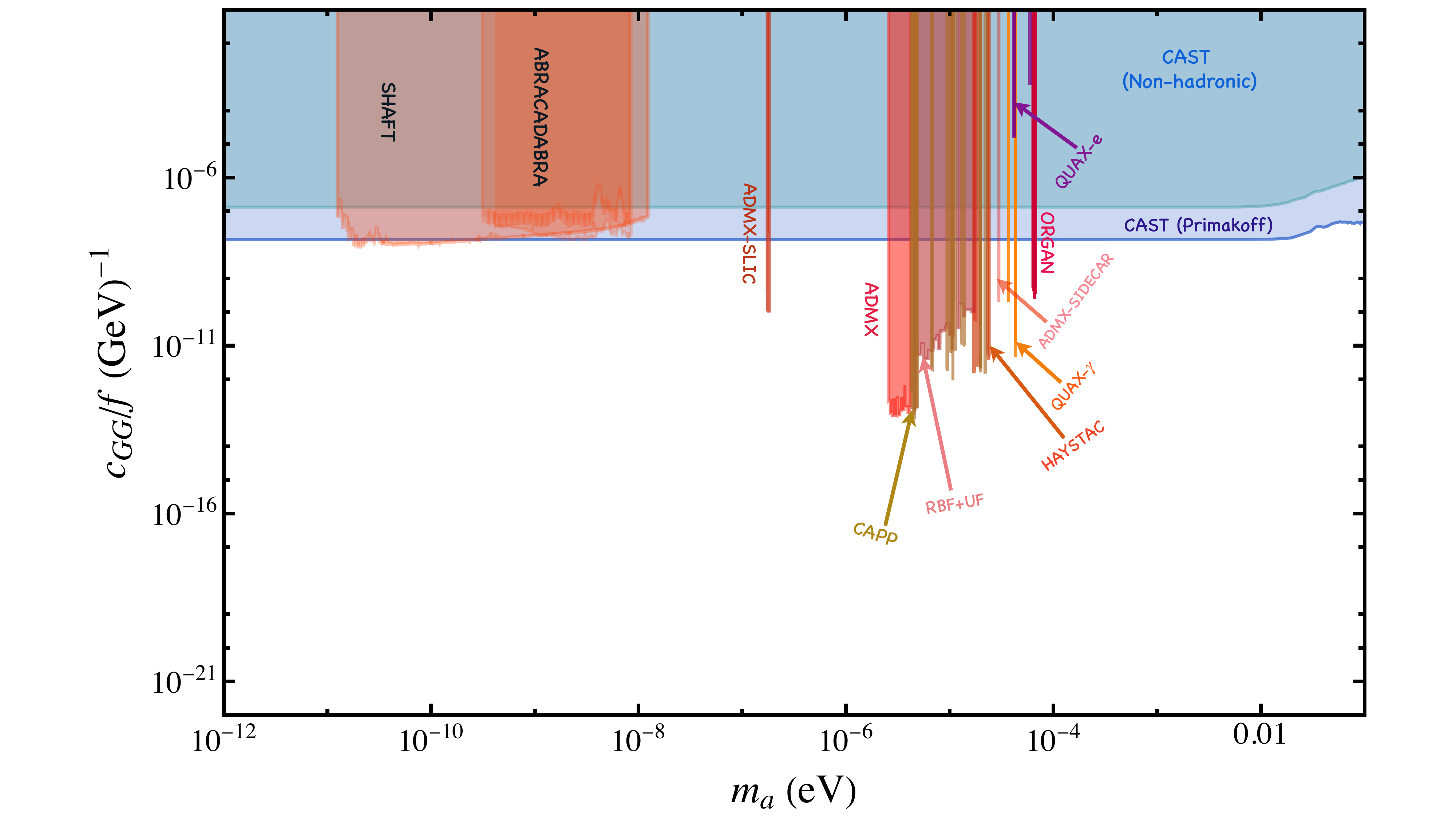}
\caption{\small Limits on the ALP-gluon coupling at the UV scale from existing {\it haloscopes} in different shades of red and {\it helioscopes} in blue and teal.}\label{fig:halo-current}
\end{figure}

In Figure~\ref{fig:halo-current}, we show the sensitivities of different cavity haloscopes on the UV scale coupling $c_{GG}/f$ as a function of the ALP mass $m_a$ induced by the linear ALP-diphoton interaction in \eqref{LlowE}. Resonant searches have the best coverage around $m_a \sim \mathcal{O}\,(\mu{\rm eV})$ which can be probed with cavities with a length of roughly a meter and a volume $\sim$ 100~litre. Searches for smaller ALP masses would require larger cavities because the volume scales as $V \propto\, 1/m_a^3$, whereas for probing higher masses, the reduced cavity volume significantly affects the signal power and consequently the scan rate. 
Among the microwave cavity searches, the best limits are obtained from {\fontfamily{qcr}\selectfont ADMX} (Axion Dark Matter Experiment) which over the course of four previous runs~\cite{ADMX:2001dbg,ADMX:2018gho,ADMX:2019uok,ADMX:2021nhd}, covers ALP mass range $m_a \approx 2.6\, - \,4.2\, \mu{\rm eV}$. For higher masses, the most sensitive search is by the
{\fontfamily{qcr}\selectfont CAPP} (Center for Axion and Precision Physics Research) experiment, covering a slightly higher mass range from almost $m_a\approx 4.24-22\, \mu{\rm eV}$ in several narrow discontinuous resonant bands that correspond to the results quoted from the previous runs~\cite{Jeong:2020cwz,Yoon:2022gzp,Kim:2022hmg,Yi:2022fmn,Kim:2023vpo,CAPP:2024dtx}. In the mass ranges $m_a \approx 4.55-10.79\, \mu{\rm eV}$ and $m_a \approx 11.96-17.35\, \mu{\rm eV}$ the experiments  {\fontfamily{qcr}\selectfont RBF-UF}
~\cite{Hagmann:1990tj,DePanfilis:1987dk} set the strongest constraint and the results from phase-I of {\fontfamily{qcr}\selectfont HAYSTAC} (Haloscope At Yale Sensitive To Axion Cold dark matter) cover the mass range $m_a \approx 23.16-23.98$ $\mu{\rm eV}$, making it the most sensitive microwave cavity haloscope experiment for ALP masses $m_a> 20\, \mu{\rm eV}$ range. HAYSTAC uses the Josephson Parametric Amplifier to minimize system noise essential for stability in the high frequency/mass range. The phase-II results~\cite{HAYSTAC:2020kwv,HAYSTAC:2023cam} cover the ranges $m_a\approx 16.95-17.27\, \mu{\rm eV}$ and $m_a \approx 18.44-18.89 \, \mu{\rm eV}$.\\

Dedicated experiments such as {\fontfamily{qcr}\selectfont ADMX-SIDECAR} and {\fontfamily{qcr}\selectfont ORGAN} have been proposed to probe the higher frequencies and ALP masses. {\fontfamily{qcr}\selectfont ADMX-SIDECAR} deals with the issue of small cavity volume at high frequencies by using a miniature resonant cavity and a piezoelectric actuator that helps tune to higher cavity modes. As a result, data are measured in two modes - ${\rm TM_{010}}$ and ${\rm TM_{020}}$ (Transverse Magnetic Modes). The latter affords sensitivity to higher frequencies with a much higher effective cavity volume than compared to ${\rm TM_{010}}$. The latest run reports the coverage of three widely spaced mass range, {\it ie,} $m_a \approx 17.36-17.57\mu{\rm eV}\, ({\rm TM_{010}}),\,m_a \approx 21-23.93\,\mu{\rm eV}\, ({\rm TM_{010}})\,  {\rm and}\, m_a \approx 29.64-29.76\,\mu{\rm eV} ({\rm TM_{020}}) $ \cite{ADMX:2018ogs}. On the other hand, {\fontfamily{qcr}\selectfont ORGAN} uses a copper resonant cavity sensitive to ${\rm TM_{020}}$. Higher modes correspond to a smaller cavity form factor but a larger volume can compensate for that, keeping the effective cavity volume $V_\text{eff}=V\,C$ constant~\cite{McAllister:2017lkb}
 and at the same time maintaining a higher scan rate because of the sizeable volume. Phase 1a run~\cite{Quiskamp:2022pks} probes the ALP mass range $m_a \approx 63.2-67.1\, \mu{\rm eV}$ while the latest phase 1b run~\cite{Quiskamp:2023ehr} covers $m_a \approx 107.42-111.93\, \mu{\rm eV}$, representing the most sensitive ALP haloscope measurement to date in the $m_a \sim 100\, \mu{\rm eV}$ range.

For lower ALP masses, $ie$,  $m_a \lesssim 10^{-5}\, {\rm eV}$, there are limits from {\fontfamily{qcr}\selectfont ADMX-SLIC}, {\fontfamily{qcr}\selectfont SHAFT} and {\fontfamily{qcr}\selectfont ABRACADABRA} which use tuneable LC circuits instead of microwave cavities to avoid dealing with extremely large-size cavities at low frequencies or ALP masses. While {\fontfamily{qcr}\selectfont ADMX-SLIC} uses a resonant LC circuit with piezoelectric-driven capacitive tuning and as per the latest run~\cite{Crisosto:2019fcj} covers several narrow resonant bands in the ALP mass window $m_a \approx  0.17-0.18\, \mu{\rm eV}$, {\fontfamily{qcr}\selectfont SHAFT} and {\fontfamily{qcr}\selectfont ABRACADABRA} are based on broadband configurations and use toroidal magnets. Although the sensitivities of these broadband searches are low compared to the cavity resonance searches, {\fontfamily{qcr}\selectfont ABRACADABRA} Run 2~\cite{Salemi:2021gck} provides competitive limits for $m_a \approx 0.31-8.3\, {\rm neV}$ and {\fontfamily{qcr}\selectfont SHAFT}~\cite{Gramolin:2020ict} bounds correspond to an even wider mass range $m_a \approx 0.012 - 12.3\, {\rm neV}$. New techniques to enhance the axion conversion rate could improve these constraints and target a wider range of ALP masses~\cite{Daw:2018qwb,Liu:2018icu}.

Conventionally, cavity haloscopes probe the ALP-diphoton coupling as the signal rate is proportional to the power originating from the ALP conversion into photons through Primakoff production inside the cavity. However, there are recent developments with a ferromagnetic haloscope {\fontfamily{qcr}\selectfont QUAX}~\cite{QUAX:2020adt} where ALP dark matter detection is based on the principle of magnetic resonances, e.g. electron spin resonances (ESR) or ferromagnetic resonances (FMR), induced by the ALP dark matter cloud acting as an effective radio frequency magnetic field $\vec{B}_a$ on the electron spins in a ferromagnetic material. In terms of the linear ALP-electron coupling in the low-energy Lagrangian of \eqref{LlowE} one can write the relevant interaction as
\begin{align}
-2\mu_B \frac{c_{ee}}{2e}\frac{m_e}{f}\vec \sigma\cdot a \equiv -2\mu_B \vec \sigma\cdot  \vec B_a\,,
\end{align}
where $\vec{\sigma}$ is proportional to the electron spin vector, $e$ is the unit charge and $\mu_B$ denotes the Bohr  magneton~\cite{Barbieri:1985cp, QUAX:2020adt}. The effective magnetic field $\vec{B}_a$ is a function of the ALP mass and coupling. It induces a variable magnetisation $M_a\propto\,  \mu_B B_a$ in the transverse plane in the sample, which is also magnetised in the direction of a uniform external static magnetic field $\vec{B}_0\, \perp \vec{B}_a$. Due to the external magnetic field $\vec{B}_0$, the material absorbs electromagnetic radiation at Larmour frequency ($\nu_L$). Dark matter detection occurs when $\nu_L$ matches the ALP frequency $\nu_a$ and power is deposited in a sample due to resonant ALP conversion to magnons as per the relation 
\begin{align}
P_a= \vec{B}_a\cdot\frac{\vec{dM}}{dt}V_S\, \propto\, B_a^2\, \nu_a V_S\,,
\end{align}
where $V_S$ is the size of the sample~\cite{QUAX:slide}.
{\fontfamily{qcr}\selectfont QUAX} 
measurements of the linear axion-electron coupling~\footnote{The QUAX cavity has also been used to detect axion-photon conversion through the standard Primakoff process upon removing the magnetic material from the cavity. {\fontfamily{qcr}\selectfont QUAX-e$\gamma$} can probe axion-diphoton coupling around $m_a\sim~43\, \mu{\rm eV}$~\cite{Alesini:2020vny}. The high $Q$ factor ensures higher stability and provides best sensitivity in the relevant ALP mass range.} are sensitive to the ALP mass window $m_a\approx 42.4-43.1\, \mu{\rm eV}$~\cite{QUAX:2020adt}. In Figure~\ref{fig:halo-current}, we show the bound obtained by {\fontfamily{qcr}\selectfont QUAX} in terms of limits on $c_{GG}$ at the UV scale, which leads to a weaker constraint than other haloscopes due to a $\mathcal{O}(10^{-3})$ suppression as per the running relations in \eqref{eq:running1}.

\subsubsection{Haloscope Searches for quadratic ALP interactions}

The presence of the quadratic ALP-photon vertex implies that inside the cavity there is also a finite probability for resonant double ALP conversion via their interactions with an external electromagnetic field 
\begin{align}
C_{\gamma}\, \frac{a^2}{4f^2} F_{\mu\nu}F^{\mu\nu} = C_{\gamma}\, \frac{a^2}{2f^2} (E^2-B^2)
\end{align}
The signal power in this case can be written as 
\begin{align}
P_{aa\to\gamma} &\propto
     \left(\frac{C_{\gamma}}{f^2} \frac{\rho_\text{DM}}{m_a}\right)^2\big( B_0^2+E_0^2) V C_\phi\, {\rm min}(Q_L,Q_a)\\
    &\propto \frac{\rho_\text{DM}}{m_a f^2} \frac{C_\phi}{C}\,P(a\to\gamma)\,.\label{eq:Paagamma}
\end{align}
Haloscopes looking for the Primakoff effect typically only have a uniform external magnetic field and $E_0=0$, and the cavity is optimised for cavity modes induced by the interaction in \eqref{eq:cav1}. As a result, the form factor measuring the coupling strength of the external electric field $\vec E_0$ or magnetic field $\vec B_0$ with a cavity eigenmode is of order one, e.g. $C=0.4$ for the dominant mode ${\rm TM_{010}}$ in ADMX. In fact, for a perfectly uniform external magnetic field along the length of a cylindrical cavity, the form factor for the quadratic couplings, $C_\phi=0$. Fringing effects induce a remnant sensitivity with $C_\phi= 10^{-12}$ (ADMX and ADMX-SIDECAR) and $C_\phi = 2\cdot10^{-9}$ (ORGAN). Limits from ADMX and ORGAN have been recast for a quadratically coupled scalar in ~\cite{Flambaum:2022zuq, McAllister:2022ibe}, respectively. Together with the additional suppression factors in \eqref{eq:Paagamma} results in much weaker constraints for this quadratic ALP conversion\footnote{Quadratic ALP interactions also induce temporal variations in the fundamental constants which gives rise to fractional changes in the Bohr radius and ultimately in the length of solid objects such as a haloscope cavity. Ideally, this length variation change should give rise to a frequency shift, implying a broadening of the haloscope's power spectrum. However, by a rough estimate, we found this fractional frequency shift negligibly small in our parameter range of interest ($e.g:\, \frac{\Delta\omega}{\omega}\sim 10^{-35}\ {\rm for}\ m_a \sim 40\ \mu{\rm eV}$ and $c_{\gamma\gamma}/f\sim 10^{-10}\ {\rm GeV^{-1}}$), compared even to the broadening due to dark matter velocity dispersion in the halo which is $\sim 10^{-6}$~\cite{OHare:2017yze}. Therefore the corresponding $Q$ factor does not enter in the signal power computation \eqref{eq:halo1}.}. 

\begin{figure}[t!]
 \centering
 \includegraphics[width=\textwidth]{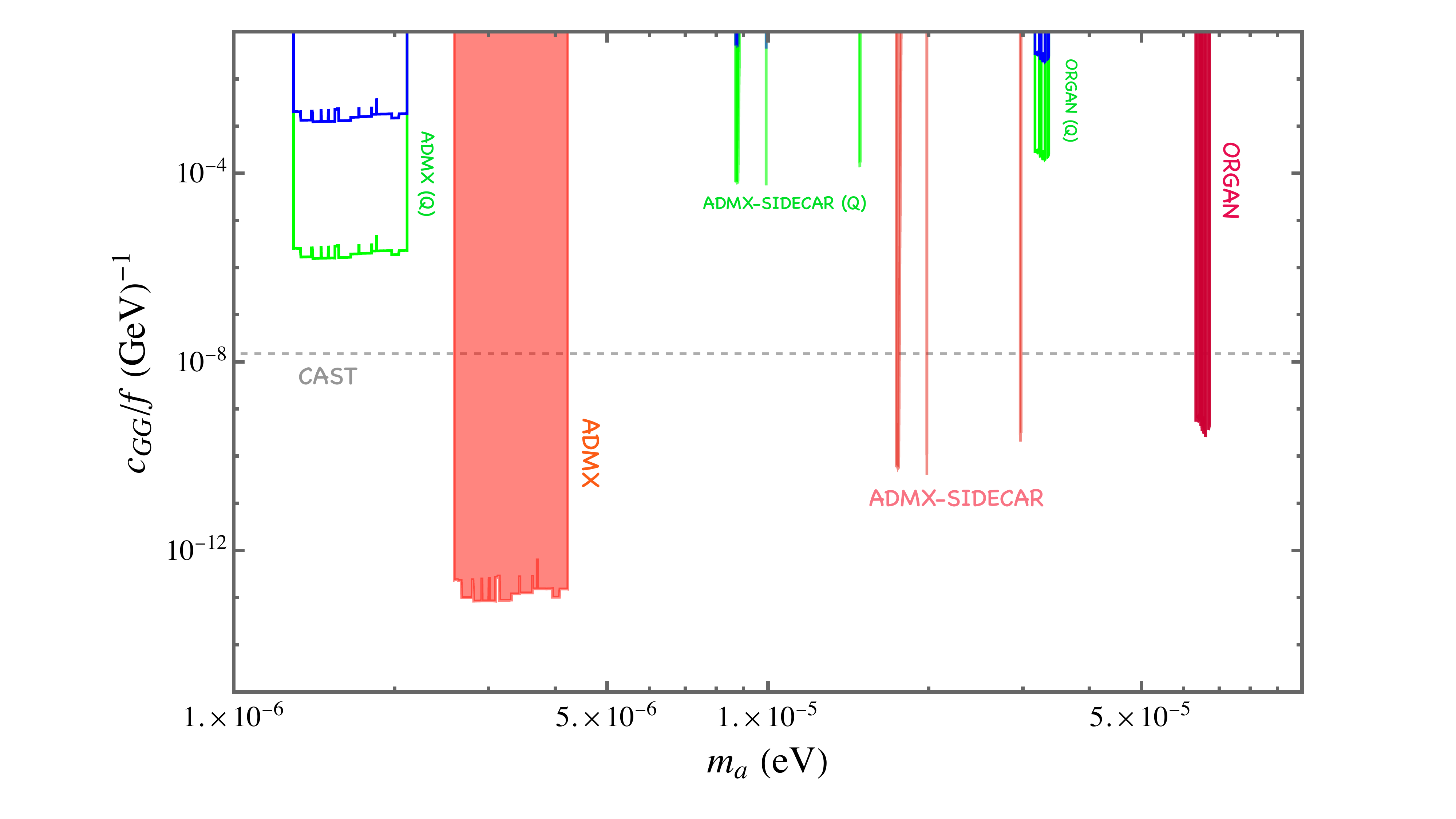}
\caption{\small Comparison of {\it haloscope} limits on the ALP-gluon coupling at the UV scale due to linear (in red) and quadratic (in green and blue) ALP-photon conversion. The green lines assume the form factor ratio $C_\phi/C = 1$
\label{fig:halo-LinQuad} whereas the blue lines incorporate $C$ and $C_\phi$ values taken from current experiments~\cite{Flambaum:2022zuq,McAllister:2022ibe}.}
\end{figure}

In Figure~\ref{fig:halo-LinQuad}, we show in a comparison plot some haloscope sensitivities in the $c_{GG}/f$ vs $m_a$ plane originating from linear and quadratic ALP couplings. In shades of red, the limits due to the linear couplings correspond to ADMX, ADMX-SIDECAR and ORGAN bounds in Figure~\ref{fig:halo-current} and the limits on the quadratic couplings have been recast using \eqref{eq:Paagamma}. The blue lines show the rescaled sensitivity for the quadratic ALP conversion using the values of $C_\phi$ and $C$ taken from \cite{Flambaum:2022zuq,McAllister:2022ibe}. Even though they are substantially weaker they can probe a new parameter space since the resonance occurs at $\omega_{\rm quad}=\omega_{\rm lin}/2$. Form factors are mode specific and in the current setups, $C_\phi$ values are tuned to very small values in order to maximise the linear ALP conversion when operating in a single mode. However, there are recent proposals for haloscopes designed to use multiple modes. Keeping in mind these possibilities, we have included the green lines in Figure~\ref{fig:halo-LinQuad} which correspond to $C_\phi/C=1$.

\subsubsection{Future Haloscopes}

\begin{figure}[t!]
 \centering
 \includegraphics[width=1\textwidth]{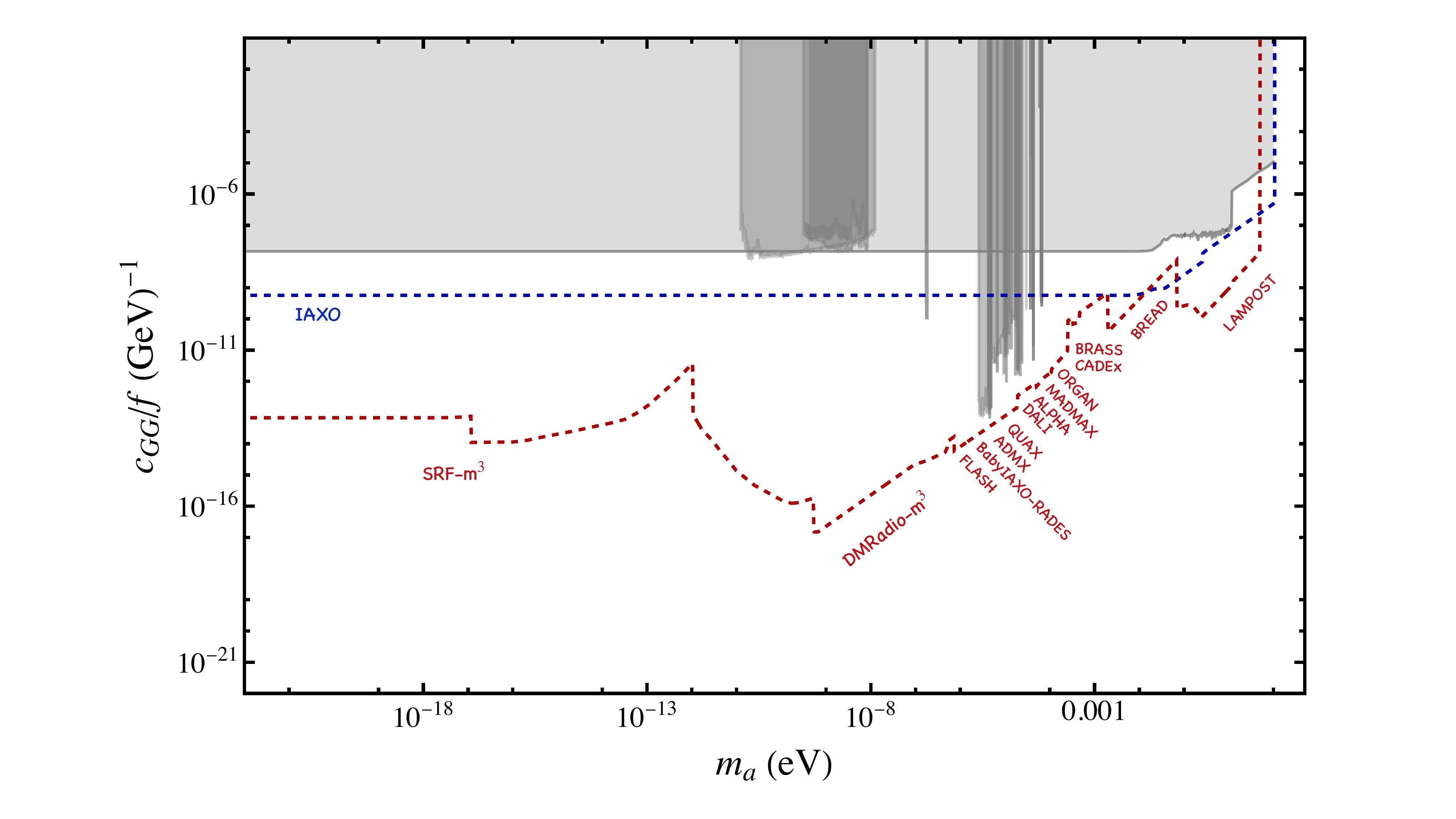}
\caption{\small Projected limits on the ALP-gluon coupling at the UV scale from future {\it haloscopes} (in red dashed line) and {\it helioscopes} (in blue dashed line). The existing constraints are shown in gray.
\label{fig:halo-future}}
\end{figure}

There are several future haloscopes proposed in near future that would substantially improve the existing limits. We show the projections from some of these experiments in red dashed line in Figure~\ref{fig:halo-future}. The projected limits cover a huge ALP mass range from $10^{-22}\ {\rm eV}$ to almost $\sim\mathcal{O}(1)$ eV. The challenges of resonant cavities operating at high frequencies with diminishing volume can be resolved by introducing dielectric haloscopes like {\fontfamily{qcr}\selectfont MADMAX}~\cite{Li:2021mep,Majorovits:2023kpz} and {\fontfamily{qcr}\selectfont LAMPOST}~\cite{Chiles:2021gxk}, which are expected to probe up to a few orders of mass above the $\mu{\rm eV}$ scale, the current operating range for the existing haloscopes. {\fontfamily{qcr}\selectfont BREAD}~\cite{BREAD:2021tpx} is expected to cover across THz frequencies/meV mass range with broadband searches. However, the most notable improvement in this category is from the experiments proposed for the lighter mass range, covering down to many orders of mass below the current coverage. ${\rm SRF{\text -}m^3}$~\cite{Berlin:2019ahk}, operating via photon frequency conversion, falls under this category. A superconducting radio frequency cavity with a high $Q$ factor is used where instead of the individual cavity mode frequencies, it is the frequency difference between the two modes that is tuned to resonance with the ALP field. ${\rm SRF{\text -}m^3}$ is projected to probe ALP masses as low as ${m_a=\rm 10^{-14}\ eV}$, whereas {\fontfamily{qcr}\selectfont DarkSRF} collaboration has recently proposed to use the same methodology in a broadband search~\cite{Berlin:2020vrk} where it would be possible to cover dark matter mass down to ${\rm 10^{-22}\ eV}$. LC resonant enhancement provides another method to avoid the problem of implausibly large cavities at low masses. Experiments like ADMX-SLIC. {\fontfamily{qcr}\selectfont DMRADIO}~\cite{Rapidis:2022gti,Adams:2022pbo} are expected to improve this technique and extend the coverage around the neV mass range with sensitivities several orders of magnitude better than ADMX-SLIC.

\subsubsection{Helioscope Searches } 

Axions are produced in abundance in the stellar cores, most importantly, in the Sun through various reactions depending on whether or not they have hadronic interactions. Solar axions that are produced are then converted back to $X$-ray photons in the electromagnetic field of the earth-based helioscopes~\cite{Irastorza:2011gs}. Similar to haloscopes, helioscopes are sensitive to the linear ALP couplings, however, unlike haloscopes, helioscopes do not assume ALPs to be the dark matter. The number of signal events, $ie,$ the total number of photons $N_\gamma$ from ALP conversion over an energy range is obtained by the following relation~\cite{Barth:2013sma}
\begin{align}
   N_\gamma=\int^{\omega_f}_{\omega_i} d\omega \left(\frac{d\Phi_a}{d\omega}\right)P_{a\to\gamma}\, \mathcal{S}\, \epsilon\, t 
\end{align}
where $\frac{d\Phi_a}{d\omega}$ is the differential ALP flux, the detector parameters $\mathcal{S}$, t and $\epsilon$ denote the surface detection area perpendicular to the axion flux, exposure time and the detection efficiency respectively. The ALP-photon conversion probability for a transverse homogeneous magnetic field $B_0= |\vec B|$ over a distance $L$ scales with the linear axion-photon coupling as
\begin{align}
    P_{a\to\gamma}=\frac{\alpha^2}{4\pi^2}\left(\frac{c_{\gamma\gamma}^\text{eff}}{f}\right)^2B_0^2L^2 {\rm sinc}^2\left(\frac{qL}{2}\right)
\end{align}
where ${\rm sinc}=\sin x/x$ and $q=m_a^2/(2\omega)$ denotes the momentum transfer. One needs $q \ll 1/L$ to ensure coherent conversion over the entire length. The signal therefore scales with the low-energy ALP couplings as $  P_{a\to\gamma}\, \propto\, c_{\gamma\gamma}^4$ for Primakoff production and $  P_{a\to\gamma}\, \propto\, c_{ee}^2 c_{\gamma\gamma}^2$ for non-hadronic production such as Compton scattering, electron-electron/ion bremsstrahlung~\cite{Barth:2013sma}.  In Figure~\ref{fig:halo-current}, we show the sensitivities from {\fontfamily{qcr}\selectfont CAST}~\cite{CAST:2017uph} where the light blue shaded region corresponds to primary ALP production through Primakoff mechanism where the signal event depends solely on the ALP-photon coupling. The teal line, on the other hand, corresponds to non-hadronic ALP production~\cite{Barth:2013sma}, for which the ALP-photon interaction is subdominant compared to the ALP-electron interaction, so it does not appear in the ALP production but does contribute to the ALP-photon conversion. In Figure~\ref{fig:halo-current}, we 
show the corresponding limits from~\cite{Barth:2013sma} in terms of the UV scale coupling $c_{GG}/f$ vs. $m_a$. The limits are practically independent of the ALP mass as long as the ALP-photon conversion is coherent, $ie,$ up to $m_a^2\le 4\omega/L$ where $L$ denotes the length of the helioscope magnet. Towards larger ALP masses, the momentum transfer becomes large and the conversion is no longer coherent. The conversion probability becomes suppressed by a factor of $(\sim 4\omega/m_a^2L)^2$ and this leads to a degradation in the limits around $m_a\gtrsim 10\ {\rm meV}$.

{\fontfamily{qcr}\selectfont IAXO}~\cite{Giannotti:2016drd}, a future helioscope proposed to be built with magnets specially designed for maximum sensitivity, is expected to improve the existing CAST limits by several orders of magnitude. We show the projections in blue dashed line in Figure~\ref{fig:halo-future}. Similarly, future quantum-sensor assisted light-shining through a wall experiments such as ALPSII are projected to deliver sensitivity beyond the CAST limit~\cite{Ortiz:2020tgs, Hallal:2020ibe}. 

\section{Bounds on ALP Dark Matter from Quantum Technology}\label{sec:probe-quad}

As already described in Section~\ref{sec:quadraticinteractions}, quadratic interactions of the ultralight axion dark matter induce coherent temporal oscillations in the fundamental constants (FC) and nuclear parameters such as the fine-structure constant, nucleon mass, electron mass, etc. The fractional time-variation of these quantities can be probed with quantum sensing technology, namely with atomic clocks, atom interferometry, mechanical resonators, etc. In this section, we will elaborate upon different sensing techniques and their respective sensitivities for our model. To demonstrate our results we again choose a model with all Wilson coefficients equal to zero in the UV, apart from $c_{GG}$.

\subsection{Quantum Clocks} 

Microwave, optical, ion, molecular and nuclear clocks, which are classed together as quantum clocks, operate by comparing the frequency ratios of different atomic, vibrational and nuclear transitions. Clock frequencies 
rely on the frequencies of spectral lines in these transitions. Therefore, a fractional change in the spectra brings in a shift in the clock frequency.  As a generic prescription for the dependence on different fundamental constants, the frequency ratio of atomic transitions in two different atomic clocks $A$ and $B$ is parametrised in terms of the fine-structure constant ($\alpha$), electron-to-proton mass ratio ($m_e/m_p$) and the ratio between quark mass ($m_q$) and QCD energy scale ($\Lambda_{\rm QCD}$)~\cite{Hees:2016gop} as
\begin{align}\label{eq:clockfreq}
    \nu_{A/B}\, \propto\,  \alpha^{k_\alpha} \left(\frac{m_e}{m_p}\right)^{k_{e}}\left(\frac{m_q}{\Lambda_{\rm QCD}}\right)^{k_q}
\end{align}
where $k_\alpha$, $k_e$ and $k_q$ are the difference between the sensitivity coefficients for the two transitions. Therefore, the fractional variation in the frequency ratio can be written as
\begin{align}
     \frac{\delta \nu_{A/B}}{\nu_{A/B}} &=k_\alpha \frac{\delta \alpha}{\alpha}+k_e\left(\frac{\delta m_e}{m_e}-\frac{\delta m_p}{m_p}\right)+k_q\left(\frac{\delta m_q}{m_q}-\frac{\delta \Lambda_{\rm QCD}}{\Lambda_{\rm QCD}}\right)
     \label{Eq:freqratio1}
\end{align}
We substitute the fractional variations in the  QCD scale $\delta \Lambda_{\rm QCD}/\Lambda_{\rm QCD}\approx\delta m_p/m_p$ and light quark mass $\delta m_q/m_q=\delta m_\pi^2/m_\pi^2$~\cite{Flambaum:2006ip, Sherrill:2023zah,Dzuba:2024src}. 
In the notations introduced previously for the above variations when induced by quadratic interactions of ALP field, \eqref{Eq:freqratio1} becomes:
\begin{align}
    \frac{\delta \nu_{A/B}}{\nu_{A/B}} &=k_\alpha\, \delta_\alpha(a) + k_e\, \delta_e(a)-\left(k_e+ k_q\right)\delta_p(a)+k_q\,\delta_\pi(a)
    \label{Eq:freqratio2}
\end{align}
where $\delta_p(a)$ is defined similar to $\delta_N(a)$ in \eqref{Eq:shift_mn} and denotes the shift in proton mass. Plugging \eqref{Eq:shift_mpi2}, \eqref{Eq:shift_mn}, \eqref{Eq:shift_alpha} and \eqref{Eq:shift_me} into \eqref{Eq:freqratio2}, one can express $\delta \nu_{A/B}/\nu_{{A/B}}$ in terms of the low-energy Lagrangian parameters and the ratio $a^2/f^2$. For an oscillating ALP dark matter field, the quadratic ALP field evolves as 
\begin{align}\label{eq:asquared}
 a^2 =\frac{2\rho_{\text DM}}{m_a^2} \cos^2\,m_a t&=\frac{\rho_{\text DM}}{m_a^2}\left(1+\,\cos \, 2m_a t \right)
\end{align}
As a result, all fractional variations of the fundamental quantities mentioned in \eqref{Eq:freqratio2} will have a constant, time-independent shift and also a time-varying part which oscillates at a frequency $\omega\simeq 2m_a$~\cite{Beadle:2023flm,Kim:2022ype}. The constant shift is discarded in many references as being unobservable due to large low-frequency stochastic background, however, recent methodologies have been proposed~\cite{Kim:2023pvt,Flambaum:2023bnw} which show that in some atomic clocks that are sensitive to low-frequency, even for a time-independent shift, the signal can be successfully extracted from the background noise of the oscillating dark matter field and therefore can provide limits for constraining ultralight dark matter parameter space.

Note that \eqref{eq:asquared} neglects the environmental effects described in Section~\ref{sec:QDM}. The profile of the ALP field close to earth \eqref{eq:DMwave2} is not well defined for a large range of the parameter space that can be accessed with quantum clocks. In our analysis, we use \eqref{eq:asquared} and express ALP quadratic couplings in the low-energy Lagrangian in terms of the ALP-gluon coupling at the UV scale, using the running and matching relations in Section~\ref{sec:theory} and subsequently set limits on the quantum sensor sensitivities in the  $c_{\rm GG}/f$ (UV scale) vs. $m_a$ parameter space. We also indicate the parameter space for which the profile close to earth is not well defined.

\subsubsection{Current limits: microwave and optical clocks}

Existing atomic clock limits have been obtained either from the {\it optical} clocks, which are based on transitions between different electronic levels, or the {\it microwave} clocks, which depend on transitions between the hyperfine substates of the atomic ground state. The clock comparison test limits, therefore, can broadly be of three types based on the frequency ratio between
\paragraph{Two microwave clocks:} Limits are obtained from the {\fontfamily{qcr}\selectfont Rb/Cs} atomic fountain clock, which compares the transition frequencies between different hyperfine levels in the two ground state atoms $^{87}$Rb and $^{133}$Cs. The frequency ratio measurements are sensitive to the variations in all three fundamental quantities in \eqref{Eq:freqratio2}~\cite{Hees:2016gop, Flambaum:2023bnw}. The measurements are also sensitive to very low frequencies corresponding to ALP mass $\sim 10^{-20}$ eV and below, due to the long time-span of the experiment. Throughout the frequency range of the experiment, the coherence time of the oscillation, $\tau_a = 2\pi/\omega/(v/c)^2$ ($v/c\sim10^{-3}$, $v$ is the virial velocity of dark matter in the galaxy) remains larger than the total timescale of measurement $T$. Although $T<< \tau_c$ maintains a coherent signal throughout the experiment, it also induces stochastic fluctuations in the dark matter amplitude~\cite{Centers:2019dyn}. Therefore, it is no longer accurate to assume the oscillation amplitude as simply $\sqrt{2\rho_{\rm DM}}/m_a$ because this will rather be a random variable with a sampling probability as determined by the distribution of the stochastic ALP field. This leads to corrections on the dark matter coupling that can vary from a factor of 2 up to 10.

\begin{figure}[t!]
 \centering
\includegraphics[width=1.2\textwidth]{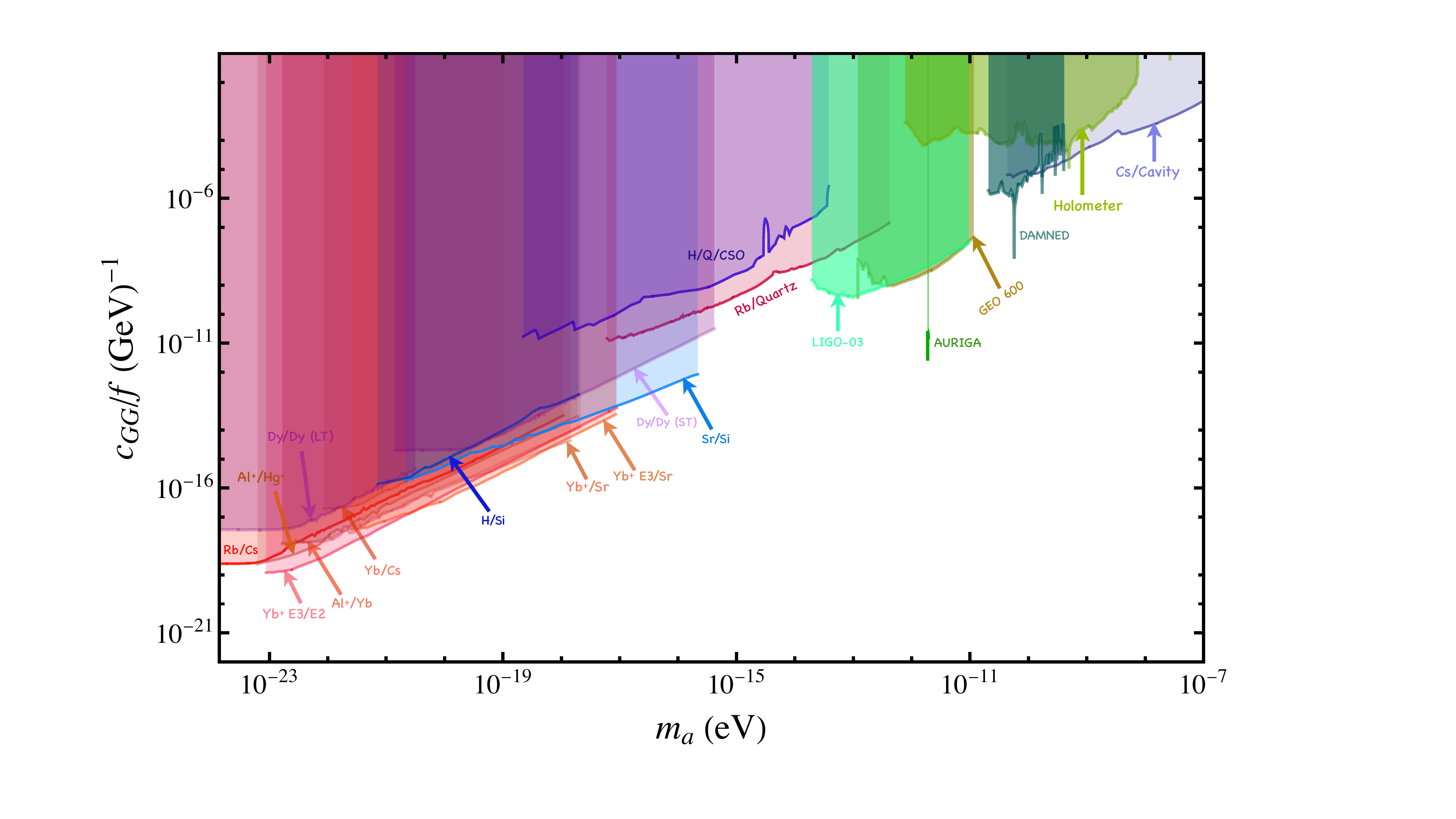}
\caption{\small Limits on the ALP-gluon coupling at the UV scale from existing {\it atomic clocks} and {\it atomic spectroscopy} (in shades of red), {\it clock/cavity comparisons} (in shades of blue), {\it optical/laser interferometers} and {\it mechanical resonators} (in shades of green).\label{fig:clocks-current limits}}
\label{current-quad}
\end{figure}

We show the bounds from {\fontfamily{qcr}\selectfont Rb/Cs} in Figure~\ref{fig:clocks-current limits} in the ${c_{GG}/f}$ ( UV scale) vs. $m_a$ plane where the coverage is for $10^{-24}\, \text{-}\, 10^{-18}\ {\rm eV}$ ALP mass\footnote{For dark matter below $\sim 10^{-22}$~eV, cosmological constraints prevent the assumption that the dark matter density of the Universe is entirely due to the ALP field because the de-Broglie wavelength of the ALP exceeds the size of a dwarf galaxy~\cite{2168507}. Therefore the low-frequency limits should be rescaled because $\rho_{\rm DM}$ should be a fraction of the total dark matter density. We show the limits assuming ALP constitutes 100\% of dark matter density.}. The constant shift in the quadratic variation in the axion field is ignored~\cite{Kim:2022ype}. The sensitivity coefficients $k_\alpha,\, k_e,\, k_q$ and the frequency comparison data are obtained from Ref.~\cite{Hees:2016gop}. The local dark matter density is assumed to be ${\rm \rho_{DM}=0.4\ GeV/cm^3}$\footnote{The bounds are shown without including the stochastic fluctuation factor appearing in dark matter amplitude as the exact correction factor is unclear for a pseudoscalar dark matter.}.

\paragraph{Two optical clocks:} An important example of this category is the frequency ratio measurements of $^{27}{\rm Al^+}$ single-ion clock and $^{171}$Yb, $^{87}$Sr optical clocks lattice clocks ({\fontfamily{qcr}\selectfont BACON})~\cite{BACON:2020ubh}. Optical transitions are only sensitive to the variation in the fine-structure constant, which implies that the sensitivity coefficients $k_e=k_q=0$. $k_\alpha$ values for ${\rm Al^{+}/Yb,\, Yb/Sr}$ and ${\rm Al^+/Hg^+}$ frequency ratio comparisons are obtained from Ref.~\cite{BACON:2020ubh}. The optical transition frequencies, being about $\sim$ 5 orders of magnitude larger than the microwave transitions, ensure better stability and the cut-off frequencies of the optical clocks are higher than the microwave clocks, which implies sensitivity to larger frequencies or dark matter masses. Bounds from BACON clocks are shown in Figure~\ref{fig:clocks-current limits} and probe a similar mass range as Rb/Cs. While $k_\alpha$ is largest for ${\rm Al^+/Hg^+}$, leading to a better sensitivity around low $m_a$, Yb/Sr provides the best limits for high masses among the three.

Other significant limits come from ${\rm Yb^+}$ ion clocks, namely ${\rm Yb^+\, E_3/E_2}$, where the frequency of the electric-octupole transition ($E_3$) is compared with the electric-dipole transition ($E_2$) of $^{171}{\rm Yb^+}$ ion, and also ${\rm Yb^+\, E_3/Sr}$, where the frequency ratio between the $E_3$ transition in $^{171}{\rm Yb^+}$ to a transition in the optical lattice clock $^{87}{\rm Sr}$ is measured. Both ${\rm Yb^+\, E_3/E_2}$ and ${\rm Yb^+\, E_3/Sr}$ have high sensitivity factors, $ie$, $k_\alpha\, \approx\, 6$~\cite{Filzinger:2023zrs}, leading to the best sensitivity among atomic clocks. Moreover, ALP couplings to quarks and gluons can lead to oscillations of the nuclear charge radius $r_N$. The variation in the so-called field shift energy $E_{\rm FS}\, \propto\, \langle r_N^2\rangle$, can be measured by comparing two electronic transitions in a heavy atom, such as $^{171}{\rm Yb^+}$~\cite{Banerjee:2023bjc}. In terms of low-energy Lagrangian parameters, this shift can be written as 
\begin{align}
    \frac{\delta \langle r_N^2\rangle}{\langle r_N^2\rangle} & \approx a\, \frac{\delta \Lambda_{\rm QCD}}{\Lambda_{\rm QCD}}+ b\, \frac{\delta m_\pi^2}{m_\pi^2} \nonumber  \approx a\, \delta_p(a)+b\, \delta_\pi(a)
    \label{Eq:chargeradius2}
\end{align}
where $a,\, b \approx$ $\mathcal{O}$ (1) in ${\rm Yb^+ E_3/E_2}$ ion clock~\cite{Banerjee:2023bjc}. We show the constraint from the oscillating nuclear charge radius in Figure~\ref{current-quad} by recasting the limits from ${\rm Yb^+ E_3/E_2}$ transition measurements. This gives the most stringent limits among quantum sensors for ALP masses $10^{-23}\, \text{-}\, 10^{-20}\ {\rm eV}$. ${\rm Yb^+\, E_3/Sr}$ takes over in the range $m_a= 10^{-20}\, \text{-}\, 10^{-17}\ {\rm eV}$.
    
\paragraph{Optical and microwave clock comparisons:} There are competitive limits in this category from frequency comparison between $^{171}$Yb optical lattice clock and $ ^{133}$Cs microwave clock. {\fontfamily{qcr}\selectfont Yb/Cs} limits are implemented in Figure~\ref{fig:clocks-current limits} with the sensitivity coefficients and the frequency ratio data taken from Ref.~\cite{Kobayashi:2022vsf}. In this case, all three sensitivity coefficients $k_\alpha, k_e $ and $k_q$ are non-zero, which makes the optical to microwave clock comparison experiments particularly sensitive to variations in $m_e$~\cite{Kobayashi:2022vsf}. For our analysis, the ALP-electron quadratic coupling is two-loop suppressed, so the $m_e$ variation remains subdominant for an ALP with only the Wilson coefficient $c_{\rm GG}/f$ in the UV. The bounds extend from $m_a=10^{-22}$ eV up to $10^{-19}$ eV.

\begin{figure}[t!]
 \centering
\includegraphics[width=\textwidth]{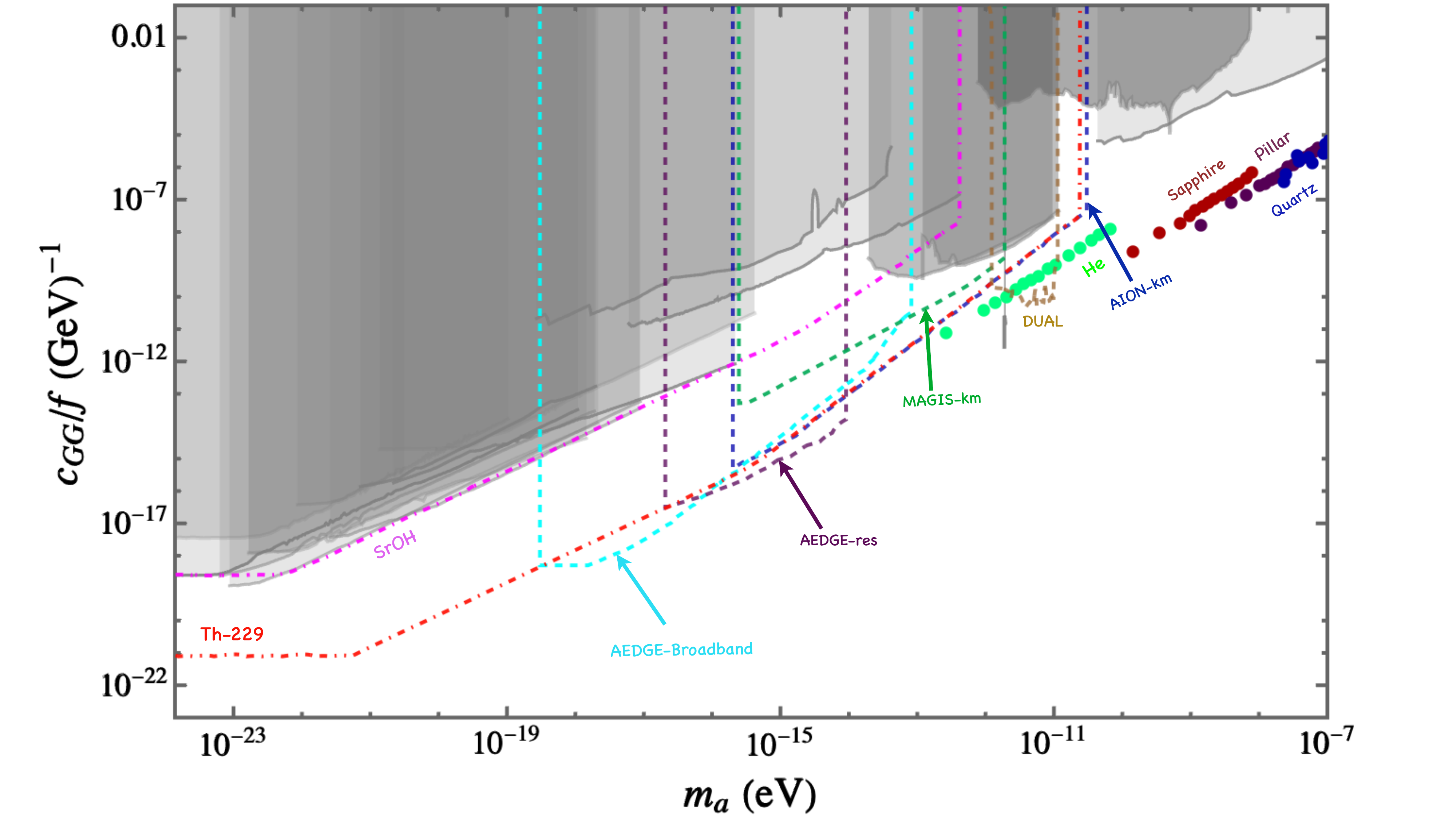}
\caption{\small Future limits on the ALP-gluon coupling at the UV scale from {\it molecular and nuclear clocks} (Th-229 and SrOH), {\it atomic interferometers} (AEDGE in broadband and resonance modes, AION-km and MAGIS-km) and {\it mechanical resonators} (DUAL, He, sapphire, micropillar and quartz BAW resonator). All the limits are sensitive to ALP quadratic couplings at low energy. Existing constraints are shown in gray.} 
\label{fig: future-quad}
\end{figure}

\subsubsection{Future projections: Nuclear and Molecular clocks}

One of the most promising future clock comparison tests is a potential nuclear clock operating on a narrow isomer transition in $^{229}{\rm Th}$. {\fontfamily{qcr}\selectfont Th-229} has an exceptionally low-energy excited isomer state with an excitation energy of a few eV, making it the only nuclear transition accessible to lasers and precision spectroscopy~\cite{Arvanitaki:2014faa}. This is the result of a cancellation between the nuclear energy shift and the electromagnetic energy shift, which means that it is particularly sensitive to new physics effects spoiling this cancellation. This corresponds to sensitivity coefficients of $k_\alpha, k_q\sim  {\mathcal O(10^4-10^5)}$~\cite{Kim:2022ype,Kim:2023pvt,Caputo:2024doz}. This increases the sensitivity of the measurements of the fundamental couplings by a few orders than the current clock comparison bounds. In Figure~\ref{fig: future-quad}, we incorporate the projected limits from Th-229 (assuming the integration time $\sim\,  10^6\ {\rm sec}$ and the averaging time $\sim 1\ \rm{sec}$~\cite{Banerjee:2020kww}) which shows that nuclear clocks could probe much smaller coupling strengths and the projections also extend to higher masses (up to $m_a\sim 10^{-11}\ {\rm eV}$) compared to current clock limits, implying better stability at high frequencies. Similarly, measurements of oscillating nuclear charge radii are projected to improve sensitivity further~\cite{Banerjee:2023bjc}.

Molecular clocks are particularly sensitive to variations in $\mu= m_p/m_e$, via vibrational and rotational transitions in diatomic and polyatomic molecules~\cite{Kozlov:2013lha,PhysRevLett.100.043202,Kozlov:2012au}. The enhanced sensitivity to this ratio is characterised by the high sensitivity coefficient associated with $\mu$ variation ($k_\mu >> 1$). Several molecular clocks have been proposed so far~\cite{Flambaum:2007wf,Kondov:2019jzq,Leung:2021zcy}, all featuring transitions between nearly-degenerate vibrational energy levels in polyatomic molecules. A significant example is the rovibrational transition in laser-cooled linear triatomic XYZ-type {\fontfamily{qcr}\selectfont strontium monohydroxide
(SrOH)} molecule~\cite{Kozyryev:2018pcp}. The proposed setup is projected to achieve $k_\mu \approx 10-10^{-3}$ when the rovibrational transitions of the $\tilde{X}(200)\longleftrightarrow\tilde{X}(03^10)$ causes excitation in the (1-30) GHz transition frequency band. In Figure~\ref{fig: future-quad}, we show the projections covering up to $m_a\sim 10^{-13}\ {\rm eV}$. The bounds correspond to the limits on $\delta\mu/\mu$~\cite{Kozyryev:2018pcp} that are translated from $\delta\omega/\omega$ sensitivity projections (SNR=1, 52 days integration time, $|k_\mu|\approx 617$) corresponding to a model that features dark matter couplings with electron, nucleon and the symmetric combination of u and d quarks, all contributing to the $\mu$ variation. The sensitivity diminishes at high frequencies and the upper limit corresponds to the Nyquist frequency.

\subsubsection{Atomic spectroscopy and Quartz oscillator limits} 

Variations of the fine-structure constant can also be probed with spectroscopic analyses. Strong limits are obtained by comparisons of transition frequencies for the two isotopes of Dy atom. In contrast with the clock-comparison test, where the transition frequencies in two different atoms are compared, here one uses a nearly degenerate pair of dysprosium isotopes of atomic masses $A= 162\ {\rm and}\ 164$~\cite{VanTilburg:2015oza}. Using spectroscopy of the radio frequency electric-dipole transition, it has been found that the energy splitting between the two isotopes corresponds to frequencies less than 2000~MHz and the splitting is extremely sensitive to the variation of $\alpha$. In contrast to the optical frequency measurements of the atom or ion clocks, which show similar sensitivity to $\alpha$ variation, the near-degeneracy of the energy levels in $^{162}{\rm Dy}$ and $^{164}{\rm Dy}$ relaxes the fractional accuracy and stability requirements of the frequency source. In Ref.~\cite{VanTilburg:2015oza}, two categories of {\fontfamily{qcr}\selectfont Dy/Dy} spectroscopic measurements are discussed: (i) long-term (LT) measurement which is based on the measurements of $\nu_{162}$ and $\nu_{164}$ averaged over 9 days and (ii) short term  (ST) measurements where the data were taken in one day over a span of 14.5 h. The LT data, being taken for longer period of time, probes lower frequency/masses and ST data provide competitive limits in the higher frequencies where the maximum analysed frequency is defined as $\omega_{\rm max}\sim 4\pi/\Delta t_{\rm min}$, where $\Delta t_{\rm min}$ is the shortest time between two successive measurements. 
In Figure~\ref{current-quad}, we show the limits from Dy/Dy measurements, which is the only complementary (but weaker) bound to Rb/Cs around $m_a\lesssim 10^{-23}\, {\rm eV}$, but it extends to higher masses compared to other clock-comparison experiments. In fact, over a small mass window  $m_a \approx 1.85\times 10^{-16}\, \text{-}\, 3.68\times 10^{-16}\ {\rm eV}$, Dy/Dy provides the best limits.\\
Frequency comparisons between ground-state hyperfine transitions in $^{87}{\rm Rb}$ and a quartz-crystal oscillator~\cite{Zhang:2022ewz} is an interesting experiment because of the high stability factor of the quartz-oscillators over a wide range of frequencies. Moreover, unlike the microwave clock comparisons, here all the sensitivity coefficients are non-zero. {\fontfamily{qcr}\selectfont Rb/quartz} probes higher frequencies than the atomic clocks and gives the best limits in the range  $m_a \approx 4.45\times10^{-16}\, \text{-}\, 2.39\times 10^{-13}\ {\rm eV}$, as shown in Figure~\ref{fig:clocks-current limits}.

\subsection{Optical Cavities and Clock-Cavity Comparisons} 

The variations of the fundamental constants due to dark matter oscillation could also induce a change in the length of solid objects such as optical cavities due to variations in Bohr radius. The fractional change in the cavity length causes a change in the frequency of the eigenmodes of the cavity, which scales as the inverse of the cavity length. Following a similar methodology as in the case of clock comparison tests, the cavity reference frequency, $\nu_c\, \propto\, \alpha\, m_e$, can be compared to the atomic transition frequencies in the clocks or other cavities in the optical/microwave domain. The most sensitive realisations of this setup are 
the frequency comparison between a Si optical cavity and a $^{87}\rm{Sr}$ optical lattice clock ({\fontfamily{qcr}\selectfont Sr/Si})~\cite{Kennedy:2020bac} and the comparison of the reference frequency of a Si cavity and an H maser ({\fontfamily{qcr}\selectfont H/Si}). While Sr/Si measurements are only sensitive to the variation in the fine-structure constant $\nu_{\rm Sr}\, \propto\, \alpha^{2.06}\, m_e$, H/Si comparisons remain sensitive to both $\alpha$ and $m_e$ variation because the hyperfine transition frequency of H maser shows a different functional dependence on $m_e$ ($\nu_H\, \propto\, \alpha^4 m_e^2$) compared to $\nu_c$\footnote{For H maser there is also a small dependence on the quark mass in its transition frequency, but it is subleading compared to the $\alpha$ and $m_e$ variation, hence neglected.}. Owing to this, H/Si limits show slightly better sensitivity than Sr/Si but cover ALP masses only up to $m_a \approx 10^{-18}$ eV. Whereas, Sr/Si operates in the optical domain with higher frequency stability and therefore provides the strongest limits in the range $m_a \approx 10^{-17}\, \text{-}\, 2\times 10^{-16}\ {\rm eV}$.

Another setup allows for the comparison of the frequency of a quartz crystal bulk acoustic wave oscillator (Q) compared with that of a H maser (H) and a cryogenic sapphire oscillator (CSO)~\cite{Campbell:2020fvq}. This measurement {\fontfamily{qcr}\selectfont H/Q/CSO} provides competitive bounds within the range $m_a\approx 10^{-16}\, \text{-}\, 10^{-14}\ {\rm eV}$, although Rb/Quartz, owing to larger sensitivity coefficients, gives better sensitivity.

Large ALP masses can be constrained by measurements of an electronic transition between two states of $^{133}{\rm Cs}$, which is resonantly excited with a laser field inside a cavity ({\fontfamily{qcr}\selectfont Cs/cavity}). The fractional variation of the difference between the atomic transition frequency ($f_{\rm {atm}}$) and the laser frequency ($f_{\rm L}$), denoted by ($\delta f/f$), is sensitive to the different dependence of the two frequencies on the fundamental constants. In the case of fast oscillations of fundamental constants, $\delta f/f$ remains sensitive to FC variations between the acoustic cut-off frequency of the cavity resonator ($\approx\, 50\ \rm{kHz}$) and the frequency corresponding to the natural linewidth of the excited state ($\approx\, 5\ \rm{MHz}$)~\cite{Antypas:2019qji}. For frequencies larger than the upper cut-off, $\delta f/f$ decreases as $1/f$. In Figure~\ref{current-quad}, we show the limits from ``Apparatus B" in ~\cite{Tretiak:2022ndx}, based on Doppler broadband spectroscopy of the $F=4\to F^\prime=3,4,5$ components of the Cs D2 line, which constrains ALPs in the mass range $m_a\approx 4.6\times 10^{-11}\, \text{-}\, 10^{-7}\ {\rm eV}$.

\subsection{Optical/laser Interferometers}

Optical interferometers are sensitive to the difference in the optical phase difference between the two interferometer arms, resulting from the variations in the dimensions of the interferometer beamsplitter. Oscillations of the fine structure constant and the electron mass cause shifts in the lattice spacing and the electronic modes of a solid, causing variations in the length $l$ and refractive index $n$ of the beamsplitter in the interferometers inside GW detectors such as GEO-600 and LIGO. In the case of {\fontfamily{qcr}\selectfont GEO-600}~\cite{Vermeulen:2021epa}, a modified Michelson's interfermeter,  these variations can be expressed in terms of fundamental constants as
\begin{align}
    \frac{\delta l}{l}&=-\left(\frac{\delta \alpha}{\alpha}+\frac{\delta m_e}{m_e}\right)\notag \\
    \frac{\delta n}{n}&=-5\times10^{-3}\left(2\, \frac{\delta \alpha}{\alpha}+\frac{\delta m_e}{m_e}\right)
\end{align}
which is valid in the limit where the mechanical resonance frequency of the beamsplitter $f_0$ is much larger than the dark matter oscillation frequency $f_0 \gg f_a$ ($f_a$ is also called the oscillator driving frequency). These variations lead to a difference in the optical path length of the two arms defined by
\begin{align}
    \delta (L_x-L_y)&=\sqrt{2}\left[\left(n-\frac{1}{2}\right)\delta l +l\, \delta n\right]
    \approx n\, l\, \left[\delta_\alpha(a)+\delta_e(a)\right]
    \label{eq:geo600}
\end{align}
where we neglect the small variation in $\delta n$.
Note however that for $f_a \gg f_0$, it is  $\delta n$ that dominates the optical path difference expression as $\delta l$ is suppressed by a factor of $f_0^2/f_a^2$. The differential strain is measured in GEO-600 as a function of frequency and \eqref{eq:geo600} can be used to set bounds on the ALP couplings because the entire optimal frequency range of the detector (100 Hz -10 kHz) remains smaller than the fundamental frequency of the longitudinal oscillation mode, which for GEO-600 is $\sim 37\  {\rm kHz}$~\cite{Grote:2019uvn}. In Figure~\ref{current-quad}, we show the exclusion limit for the ALP mass range $m_a\approx 10^{-11}\, \text{-}\, 10^{-13}\ {\rm eV}$.

The {\fontfamily{qcr}\selectfont Fermilab Holometer}~\cite{Aiello:2021wlp}, on the other hand, uses two identical, spatially separated Michelson interferometers and measures the coherent average of the cross-spectrum. The length of the interferometer arm and the separation between the two interferometers, both being much smaller than the reduced de-Broglie wavelength of dark matter over the optimal frequency range ensures the coherence of the dark matter field throughout. The cross-spectrum measurement substantially increases the signal-to-noise ratio in comparison to the single-interferometer setups. Holometer covers a higher mass range than GEO-600, due to the fundamental oscillation frequency of the beamsplitter being $\sim 226\ {\rm kHz}$. The Holometer limits in Figure~\ref{current-quad} extend from $m_a\approx 10^{-12}\, {\rm eV}$ to almost $10^{-8}\ {\rm eV}$.

In laser interferometers like {\fontfamily{qcr}\selectfont DAMNED}~\cite{Savalle:2020vgz}, a laser source is locked onto an ultrastable cavity with a locking bandwidth $\sim 100\ {\rm kHz}$ and it is unevenly distributed over the three interferometer arms of the setup, causing a de-synchronisation in the signal phase at different points in time. The oscillations of the fundamental constants cause variations in the cavity output frequency ($\omega$) and also in the fibre delay, which is given by $T=n\, L_f/c$ with $n$ and $L_f$ being the refractive index and the length of the fibre. Both these effects cause an oscillatory pattern in the signal phase between the delayed and the non-delayed signal and the resulting phase difference 
\begin{align}
\Delta \Phi(t)\, \propto\left(\frac{\delta \omega}{\omega_0}+\frac{\delta T}{T_0}\right)\,, 
\end{align}
which can be expressed in terms of variations of the fundamental constants~\cite{Savalle:2019jsb}
\begin{align}
\label{eq:damned1}
    \frac{\delta \omega(t)}{\omega_0}&\approx-(\delta_\alpha+\delta_e)\frac{\rho_\text{DM}}{m_a^2f^2}( (1+A)\cos\, (2\omega_a t)+B \sin\, (2\omega_a t))\notag\\
    \frac{\delta T(t)}{T_0}&=\frac{\delta L_f(t)}{L_{f_0}}+\frac{\delta n(t)}{n_0} 
\end{align}
where $L_{f_0}$ and $\omega_0$ are the unperturbed length and frequency of the cavity, respectively. 
The coefficients $A, B$ are negligible below resonance, on resonance $A=0, B \sim \mathcal{O}(10^4)$ and above resonance $A=-1, B=0$. The respective variations in the length and the refractive index of the fibre can be expressed as
\begin{align}
\label{eq:damned2}
    \frac{\delta L_f(t)}{L_{f_0}}& \approx -(\delta_\alpha+\delta_e)\frac{\rho_\text{DM}}{m_a^2f^2}\cos\, (2\omega_a t)\\
    \frac{\delta n(t)}{n}&\approx 10^{-2}\bigg(\frac{\delta \omega(t)}{\omega_0}-
    [2\delta \alpha+\delta_e+(\delta_e-\delta_p)/2-0.024(\delta_q-\delta_p)]\,\frac{\rho_\text{DM}}{m_a^2f^2} \cos\,(2\omega_a t)\bigg)
\end{align}
The oscillation data for the phase fluctuation $\Delta\Phi(t)$ can be used to set limits on the ALP couplings as a function of dark matter mass. The corresponding bounds are shown in Figure~\ref{current-quad}, covering the ALP mass range $m_a\approx 2\times 10^{-11}\, \text{-}\, 4\times 10^{-11}\ {\rm eV}$. Peaks occur at resonances corresponding to the Compton frequency $2\, \omega_a \approx \omega_r$, where $\omega_r$ denotes the resonant frequency of the cavity, with $\omega_r=2\pi n\, 27.62\, {\rm kHz}$ and $n=1,3,5,7$.

Oscillations of fundamental constants can also be probed with Febry-Perot interferometers like LIGO because they change the dimension and the refractive index of the beamsplitter~\cite{Gottel:2024cfj}. The methodology is similar to GEO600 due to the signal being proportional to the length variation of the beamsplitter (variation in the refractive index is subdominant), but for LIGO the sensitivity is attenuated by a factor of arm cavity finesse $\sim \mathcal{O}(100)$. There is an additional contribution to $\delta (L_x-L_y)$ from the thickness variation of the mirrors fitted on the two cavity arms. However, this is a subleading effect because $\delta(L_x-L_y)$ is proportional to the thickness difference between the mirrors in the two arms which is tiny ($\sim 80\ \mu {\rm m}$) in the current LIGO setup~\cite{Grote:2019uvn}. We include these effects from {\fontfamily{qcr}\selectfont LIGO-03} observations~\cite{Gottel:2024cfj} in Figure~\ref{current-quad} and set limits in the mass range $m_a \approx 10^{-14}\, \text{-}\, 10^{-11}\ {\rm eV}$~\footnote{LIGO-O3 also probes ultralight dark matter couplings via so-called {\it acceleration effect}~\cite{Fukusumi:2023kqd}, where the interferometer mirrors are subjected to an acceleration caused by the dark matter field gradient. FC oscillations induce variations in the mass of the atoms and that in turn causes mass variation in macroscopic objects such as mirrors. Limits are obtained for scalar dark matter~\cite{Morisaki:2018htj}, although weaker than the other laser interferometer limits discussed here. For ALP dark matter, the limits should be calculated starting from the mirrors' equation of motion, similar to \eqref{eq:accquad}. We postpone this for a later study.}.

Future {\it axion} interferometers utilising polarised light have the potential to further improve the sensitivity of interferometer searches~\cite{DeRocco:2018jwe, Liu:2018icu, Martynov:2019azm}.

\subsection{Mechanical resonators}

Similar to optical cavities, mechanical resonators are sensitive to the time variation of the mechanical strain $h(t)$ of solid objects consisting of many atoms, which originates in variations of the atom size caused by the fluctuations of the fundamental constants such as $\alpha$ and $m_e$, with
\begin{align}
\label{eq:mechres}
    h(t)=-(\delta_\alpha(a)+\delta_e(a))\,.
\end{align}
For quadratic ALP couplings that induce the FC variations above, the strain can be resonantly enhanced if one of the acoustic modes of the elastic body is tuned to twice the ALP Compton frequency ($\omega_a$). One can impose competitive bounds on the axion dark matter coupling from the frequency-dependent strain data measurement in the mechanical resonators. The cryogenic resonant-mass detector {\fontfamily{qcr}\selectfont AURIGA}~\cite{Branca:2016rez, Manley:2019vxy} provides sensitivity over a narrow bandwidth (of the instrument) 850-950 Hz, which corresponds to an ALP mass window $1.88\, \text{-}\, 1.94 \ {\rm peV}$. 

While in the case of AURIGA the bar lengths of $\mathcal{O}$(m) provides sensitivity to $\sim$ kHz resonant frequencies, future compact acoustic resonators of mm-cm scale would cover Hz-MHz range frequencies~\cite{Manley:2019vxy}. In Figure~\ref{fig: future-quad}, we include the projections from {\fontfamily{qcr}\selectfont superfluid Helium bar resonator (He)} sensitive to $m_a\approx 0.25 \ {\rm peV}\,  \text{-}\, 0.65\  {\rm neV}$, {\fontfamily{qcr}\selectfont sapphire cylinder (sapphire)} covering $m_a\approx 0.14\, \text{-}\, 7.93\ {\rm neV}$, {\fontfamily{qcr}\selectfont quartz micropillar resonator (Pillar)} probing 1.36 - 71 neV and {\fontfamily{qcr}\selectfont quartz BAW resonator (Quartz)} constraining 23-589 neV dark matter mass. These resonators measure the strain sensitivity ($\mathcal{S}$) which is related to the minimum detectable mechanical strain as
\begin{align}
    h_{\rm min}=2\sqrt{\mathcal{S}}\, (\tau_{{\rm int}}\, \tau_a)^{-\frac{1}{4}}\,,\qquad \text{for}\quad \tau_\text{int} \gg \tau_a
\end{align}
where the total runtime of the measurement $\tau_{\rm int}$ is assumed to be much larger than the coherent time for the dark matter signal $\tau_a$. Using \eqref{eq:mechres}, we obtain limits on the minimum detectable ALP coupling. The sensitivity worsens at large masses because $\tau_a$ is shorter. In Figure~\ref{fig: future-quad}, we also show projections from the resonant mass Gravitational Wave detector {\fontfamily{qcr}\selectfont DUAL}~\cite{Arvanitaki:2015iga}, which is expected to cover the ALP mass range $m_a\approx 10^{-12}\, \text{-}\, 10^{-11}\ {\rm eV}$.

\subsection{Atom Interferometers}

Oscillations in the fundamental constants cause oscillation in atomic transition frequencies~\cite{Zhao:2021tie,Buchmueller:2023nll}. In atom interferometers, sequences of coherent and single-frequency laser pulses that are resonant with the transition between an atomic ground state and a specific excited state are used to split and recombine matter waves. Atomic interferometers measure the phase shift between split atomic wave packets and detect a dark matter-induced signal phase when the period of atomic transition oscillation matches the total duration of the interferometric sequence.
The oscillation of fundamental constants generates an oscillatory component in the electronic transition frequency
\begin{align}
    \omega_A(t,x)=\omega_A+\delta \omega_A(a)
\end{align}
where $\omega_A\, \propto\,  m_e\, \alpha^{2+\xi}$ with $\xi$ is a calculable transition-specific parameter and $\xi\approx 0.06$ for the cases discussed here~\cite{Buchmueller:2023nll} and
\begin{align}
    \frac{\delta\omega_A(a)}{\omega_A}&=\delta_e(a)+(2+\xi)\,\delta_\alpha(a)\notag\\
    &\approx \left(\delta_e+(2+\xi)\delta_\alpha\right)\frac{\rho_{\rm DM}}{m_a^2 f^2}\cos\, (2\omega_a t) \equiv \overline{\omega}_A \cos\, (2\omega_a t)
\end{align}
Due to this time-dependent correction to $\omega_A$, the dark matter-induced signal is accumulated in the propagation phase of the excited state relative to the ground state. For each path segment (between time $t_1$ and $t_2$) when the transition occurs, the dark matter-induced contribution to the propagation phase is therefore
\begin{align}
    \Phi_{t_1}^{t_2}=\int^{t_2}_{t_1}\delta\omega_A(a)dt
\end{align}
The total phase difference for a single atom interferometer is thus obtained by summing over all such paths in which the atom is in the excited state. However, the sensitivity of single atom interferometers is limited by the phase noise of the laser, which can be overcome in a system of two or more interferometers, where the common laser phase noise cancels~\cite{Buchmueller:2023nll}. The total phase shift in a system constituting a pair of atom interferometers, also known as a gradiometer, is given by~\cite{Badurina:2021lwr}\footnote{Ref.~\cite{Badurina:2021lwr} discusses the phase shift expression for linear dark matter couplings where the variation or fundamental constants $\propto \cos\, (\omega_a t+\theta)$. We reinterpreted it for quadratic couplings in \eqref{eq:phaseAI}.}.
\begin{align}\
   \label{eq:phaseAI}
   \Phi_s=\frac{4\, \overline{\omega}_A}{\omega_a}\frac{\Delta r}{L}\sin\,(\omega_a nL)\sin\left(\omega_a T\right)\sin\bigl\{\omega_a\left(T+(n-1)L\right)\bigr\}
\end{align}
where $\Delta r$ and $L$ are the separation between the two atom interferometers and the baseline length respectively, $n$ denotes the number of times the atoms interact with the laser pulse while getting a large momentum transfer kick each time (therefore $n$ also denotes the number of kicks) and $T$ is the interrogation time. In this notation, the total duration of a single interferometric sequence is $2T+L$ and there are total $4n-1$ laser pulses.

There are several proposals to probe ultralight dark matter with compact gradiometers such as AION-10 operating under the assumptions $nL \ll T$ and  $m_a nL\ll 1$. Using these conditions and substituting $\omega_a\approx m_a$, \eqref{eq:phaseAI} can be further simplified as
\begin{align}
    \Phi_s=4\, \overline{\omega_a}n\Delta r \sin^2\, (m_a T)
\end{align}
As stated in the above equation, the sensitivity scales linearly with the separation between the interferometers. However, the earth-based atom interferometry proposals are limited to km-scale separation. In the long-baseline experiments, $\Delta r \approx L$.

Large-scale earth-based atom interferometers such as AION and MAGIS, are proposed to be built in the near future, each starting from a 10 m baseline with the goal to eventually realise a 1 km baseline. A longer baseline corresponds to higher sensitivity to ALP couplings. Atomic interferometers discussed here are based on the $\rm{5s^2\,^1S_0-5s5p\,^3P_0}$ optical transitions in $^{87}{\rm Sr}$. There are also proposals for a space-based interferometer AEDGE, where the spatial separation between two cold atom clouds would be $\sim 10^8\, {\rm m}$. In all cases, atomic interferometer sensitivities are limited by shot noise, whereas noise from gravity gradients dominates at low frequencies and sets the lower limit on the frequency/mass probed when it surpasses shot noise~\cite{Badurina:2021lwr}.

In Figure~\ref{fig: future-quad}, we show the projected limits for {\fontfamily{qcr}\selectfont AION-km}~\cite{Badurina:2019hst}, {\fontfamily{qcr}\selectfont MAGIS-km}~\cite{MAGIS-100:2021etm}\footnote{The 10 and 100 m baseline versions for AION and MAGIS are proposed be built sooner than their km-scale counterparts, but we have shown the maximum projected coverage from these experiments.} and AEDGE. While gravity gradient noise dictates the lower bound on frequency/mass, the upper bound corresponds to the maximum frequency
at which the dark matter signal remains coherent, ie, $\tau_a \gtrsim \Delta t$, where $\Delta t$ is the time-interval between the successive measurements. For AION and MAGIS, gravity gradient noise sets the lower bound at $m_a \approx 2\times 10^{-16}$ eV. However, AION projects a better coverage than MAGIS at higher frequencies as it extends up to $m_a \approx 3\times 10^{-11}\, {\rm eV}$, almost an order of magnitude more than the highest frequency expected to be covered by MAGIS.

The projected limits from {\fontfamily{qcr}\selectfont AEDGE}~\cite{AEDGE:2019nxb} are obtained both in broadband and resonance modes. Switching from broadband to resonant mode is possible by changing the pulse sequence used to operate the device, leading to a $Q$-fold enhancement for the resonant mode,  therefore increasing sensitivity at certain frequencies. Although a better sensitivity to larger ALP masses $m_a\approx10^{-16}-10^{-14}$ eV, can be achieved operating in resonant mode, the broadband mode of AEDGE provides the best coverage at low frequencies corresponding to ALP masses of $m_a\approx10^{-18}\, {\rm eV}$. This is largely because a higher $Q$ corresponds to a shorter interrogation time $T$, which leads to a loss of sensitivity at lower masses and shifts the best sensitivity to higher masses while operating in the resonant mode~\cite{Badurina:2021lwr, Arvanitaki:2016fyj}.

\begin{figure}[!ht]
    \centering
    \includegraphics[width=.88\textwidth]{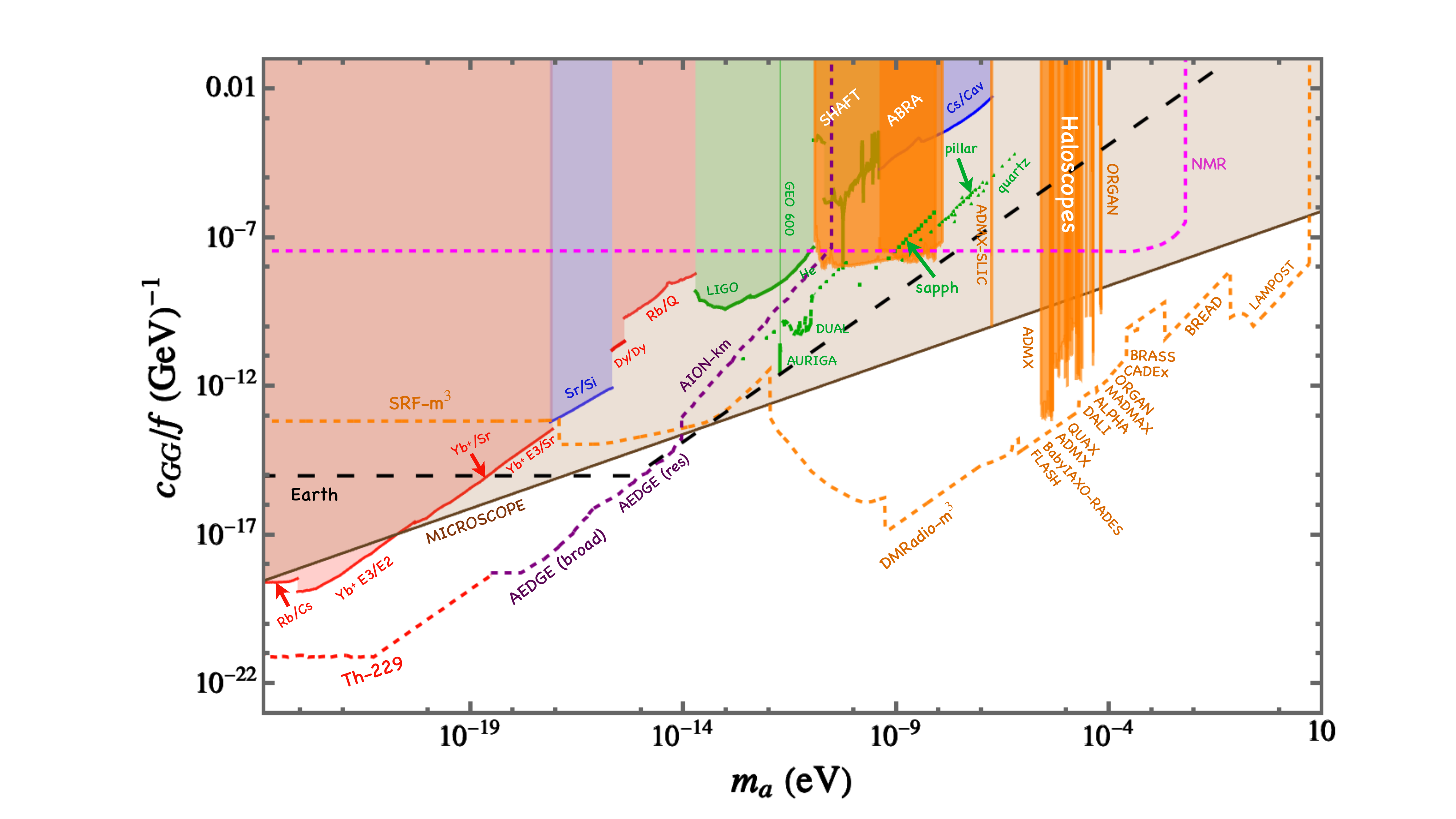}
    \caption{\small Current experimental limits (sensitivity) on ALP dark matter from - (i) quantum sensors (ii)  haloscopes and (iii)  EP-violation tests. Exclusions are shown in solid lines for {\it atomic clocks} in red, {\it clock-cavity tests} in blue, {\it optical/laser interferometers} and {\it mechanical resonators} in green, {\it haloscopes} in orange and {\it EP-violation tests} in brown. Also shown are projections in dashed lines from {\it nuclear clock} (Th-229) in red, {\it atom interferometers} (broadband and resonant modes) in purple, {\it mechanical resonators} in green, {\it haloscopes} in orange and {\it nuclear magnetic resonances} in magenta. Above the black dashed line, the small-coupling approximation breaks down, leading to an unphysical parameter space.}
    \label{fig: combDM}
\vskip-1em
\end{figure}

\begin{figure}[!ht]
    \centering
    \includegraphics[width=.88\textwidth]{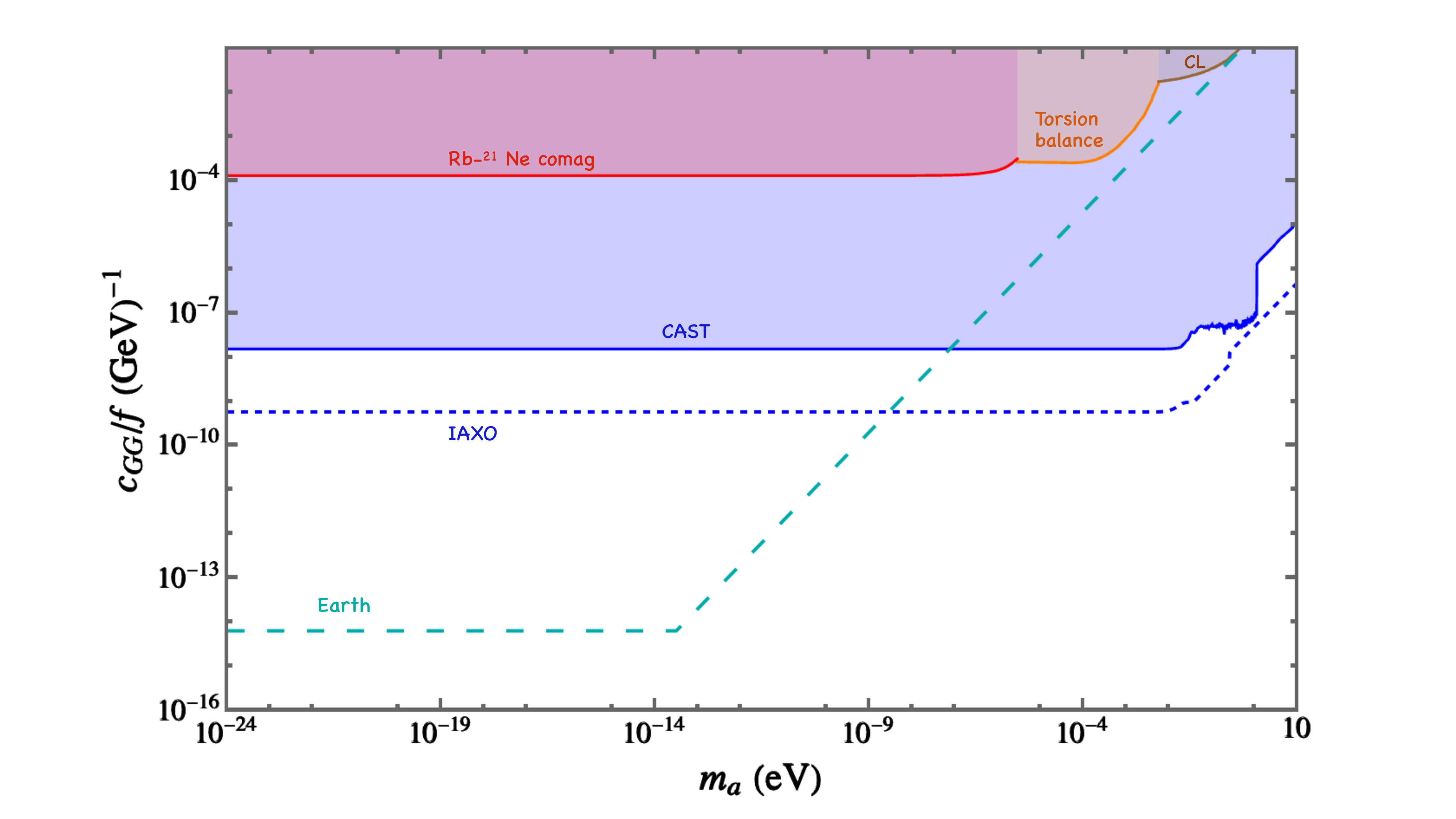}
    \caption{\small Current experimental limits (sensitivity) on the ALP-gluon coupling at the UV scale where we compare only the most stringent bounds for each mass from two categories - (i) {\it helioscopes} in blue and (ii) {\it fifth forces} in red, orange and brown solid lines and shades. The future projections are shown with dashed lines. In helioscopes, we show the current exclusions from CAST and future projections from IAXO. The current limits for fifth-forces are taken from Rb -$^{21}$Ne comagnetometer, torsion balances and Casimir-less forces. All these experiments do not require the ALP to be dark matter. The bound from non-perturbative ALP field values close to earth is shown by the green dashed contour, see ~\cite{Hook:2017psm}}  
    \label{fig: combNonDM}

\end{figure}
\section{Summary}
\label{sec:summary}

An overview of the experimental constraints on a light ALP field is shown in the $m_a - c_{GG}/f$ plane in Figure~\ref{fig: combDM} for the case that the ALP field explains the observed dark matter relic density, and in Figure~\ref{fig: combNonDM} for the case that the ALP field does not contribute to dark matter. We assume that the ALP mass is a free parameter and the UV theory has only interactions between the ALP and the $SU(3)_C$ field strength tensor. Running and matching effects generate couplings to other SM particles and induce quadratic ALP couplings in the low energy theory. Note that we assume there is no CP violating ALP coupling.

There are significant differences between the two scenarios. If light ALPs are dark matter, several effects can be observed that are not present for a light ALP field that does not contribute significantly to the dark matter relic density. The variation of fundamental constants induced by oscillations of the quadratic ALP field is the most sensitive probe for ALPs lighter than $m_a\approx 10^{-20}$ eV. These experiments probe the ALP gluon coupling via its correction to nucleon masses, photons, and electrons. Quadratic ALP interactions probe the gradient of the ALP field which leads to very strong constraints from tests of the equivalence principle for masses $m_a\approx 10^{-20}-10^{-5}$ eV. We stress however that the assumptions going into the calculation of both variations of fundamental constants and tests of the equivalence principle rely on a small-coupling approximation which breaks down for an interaction strength of $c_{GG}/f\gtrsim 10^{-15}$ GeV$^{-1}$ and $m_a\lesssim 7\times10^{8}/f\ {\rm eV^2}$, as indicated by the black dashed line in Figure~\ref{fig: combDM}. For larger couplings, the ALP dark matter field has nonlinear solutions which leads to an non-perturbative parameter space~\cite{shortpaper}. Haloscopes and helioscopes mostly probe the ALP photon coupling via the Primakoff effect in an external magnetic field. Even though haloscopes probe the linear ALP-photon interaction, they measure the local ALP field which is affected by the non-linear behaviour induced by quadratic interactions. In the absence of this effect, they are the most sensitive experiments in the mass range $m_a\approx 10^{-5}-10^{-4}$ eV. Helioscopes are insensitive to the dark matter halo and the corresponding parameter space can be considered ruled out independent of the local value of the ALP field. 

In the case where the ALP field does not contribute to the dark matter, non-perturbative ALP field values are relevant close to massive bodies for small ALP masses~\cite{Hook:2017psm}. The strongest constraints set by lab measurements for the whole mass range in this case are set by the CAST helioscope. We also show the parameter space excluded by searches for fifth forces which include the contributions from shift-invariance breaking quadratic ALP-nucleon interaction. Both probe the couplings induced by the RG running.

\section{Conclusions}\label{sec:conclusions}

We present a comprehensive analysis of the sensitivity of quantum sensors and high-precision measurements to effects from light axions or axionlike particles. Below a mass of a few electronvolt ALPs can be dark matter candidates, with the relic density produced via misalignment and the ALP field behaving like a classical (pseudo)scalar background field. We compute the most promising experimental strategy for light ALP searches and stress the complementarity of experiments that can distinguish whether ALPs make up a substantial component of dark matter. 

To this end, we go beyond the existing work on quantum sensor searches for ALPs in several important ways. First, we take effects from renormalisation group running and matching into account by expressing the low energy couplings of the ALP with SM fields in terms of the coefficients of the UV theory, which allows to compare the sensitivity of experiments at different scales. An important effect is the ALP coupling to electrons that is generated via RG effects present even if the ALP only couples to gluons in the UV. We further use the ALP interactions in the chiral Lagrangian to compute interactions quadratic in the ALP field. We also express the `dilatonic charges' in terms of the ALP couplings in the chiral Lagrangian. These interactions are nominally subleading with respect to linear ALP interactions, but induce effects that can be more relevant due to the experimental sensitivity, for example, time-dependent fundamental constants and spin-independent fifth forces. We find a clear hierarchy in quadratic ALP couplings due to the derivative nature of ALP couplings, which always leads to a suppression of quadratic ALP-electron couplings. 
The solution of the equations of motion for quadratically coupled fields in the presence of a massive body such as a planet has a position-dependent mass term which implies an non-perturbative field values for ALPs for $c_{GG}/f \gtrsim 10^{-15}$ GeV$^{-1}$ and $m_a\lesssim 7\times10^{8}/f\ {\rm eV^2}$, an effect equivalent to the one observed for ALPs that are not dark matter in~\cite{Hook:2017psm}.

Even though our formalism allows to consider any combination of ALP couplings in the UV, we use a scenario with a single ALP coupling to gluons in the UV in order to illustrate our results.
We compare constraints from the fifth forces induced by linear and quadratic ALP interactions and emphasize how quadratic interactions can lead to better sensitivity since linear ALP couplings only induce spin-dependent forces. The sensitivity of cavity searches with haloscopes and helioscopes is also significantly modified by taking into account quadratic ALP interactions. Besides a resonance induced by the conversion of a single pseudoscalar in the external magnetic field, cavities are also sensitive to the scalar interaction of two ALP fields which put constraints in a different mass range. 
Quantum clocks are one of the most sensitive probes of very light ALPs by probing variations in fundamental constants. We compare the sensitivity of optical, microwave, future nuclear, and molecular clocks with cavity searches. For low masses, clocks are the most sensitive experiments, whereas tests of the equivalence principle and haloscopes are more sensitive for higher masses. For intermediate masses, mechanical resonators are competitive with haloscopes. 

Laser and atomic interferometers provide new ways to search for ALPs. In the case of laser interferometers, the effect of the ALP background field is to cause shifts in the length of the beamsplitter, whereas for atomic interferometers it induces a phase shift in the wavefunction. Both effects are induced by quadratic ALP couplings. Together with potential future measurements with a nuclear Th-229 clock, space-based interferometers like AEDGE are capable to significantly improve sensitivity. For lower masses, higher frequency LC-oscillators like DMRADIO have the potential to significantly improve over current sensitivities. In the case where the ALP is unrelated to dark matter, the strongest constraints are set by helioscopes and could be improved by an order of magnitude once IAXO is operational.

There are several directions for future work. A full solution of the field equations close to massive bodies, taking into account the dark matter boundary conditions and the full ALP potential, is necessary to compute reliable theory predictions for the non-perturbative parameter space. Different UV models, in particular models with photon and lepton couplings, would substantially change the hierarchy of experimental limits and projections. Finally, taking into account CP violation in the Standard Model gives rise to ALP-induced neutron dipole moments, spin precession, and long-range forces from a single ALP exchange, which could considerably affect the allowed parameter space.

\section{Acknowledgements}\label{sec:acknowledgements}
MB thanks Joerg Jaeckel and Michael Spannowsky for useful discussions. We further thank Itai Bloch, Quentin Bonnefoy, Kai Bartnick, Sebastian Ellis, Gilad Perez, Inbar Savoray, Javi Serra, Konstantin Springmann, Stefan Stelzl and Andreas Weiler for useful comments and in particular for pointing out work on non-perturbative axion field configurations in the context of astrophysical observables.  MB and SC acknowledge support from the UKRI Future Leader Fellowship DarkMAP.

\appendix

\section{Connection to the UV theory}
\label{app:running}

In order to evaluate the ALP couplings in the low-energy theory considered here we have to take into account renormalisation group effects~\cite{Chala:2020wvs, Bauer:2020jbp,Bauer:2021mvw}. We consider flavor diagonal ALP coulings to SM fermions at the UV scale $\bm c_Q = c_Q \mathbbm{1}, \bm c_u = c_u \mathbbm{1}, \bm c_d = c_d \mathbbm{1}, \bm c_L = c_L \mathbbm{1}, \bm c_e = c_e \mathbbm{1}$, so that the UV theory has 8 independent ALP couplings to SM particles. Fermion couplings enter the RG equations with a Yukawa factor and we set all fermion Yukawa couplings to zero apart from the top Yukawa $y_t\sim 1$. RG running of the ALP couplings to vector bosons between the UV scale and the electroweak scale is then described by the equations
\begin{align}\label{RGEs2l}
   \frac{d\,\tilde c_{GG}}{d\ln\mu}
&= \frac{y_t^2}{8\pi^2}\,c_{tt}
    + \frac{6\alpha_s^2}{\pi^2}\,\tilde c_{GG} 
    + \frac{27\alpha_2^2}{16\pi^2}\,\tilde c_{WW}
    + \frac{11\alpha_1^2}{16\pi^2}\,\tilde c_{BB} \,, \notag\\
   \frac{d\,\tilde c_{WW}}{d\ln\mu}
   &=\frac{3y_t^2}{32\pi^2}\,c_{tt} 
    + \frac{9\alpha_s^2}{2\pi^2}\,\tilde c_{GG} 
    + \frac{27\alpha_2^2}{8\pi^2}\,\tilde c_{WW}
    + \frac{3\alpha_1^2}{8\pi^2}\,\tilde c_{BB} \,, \notag\\
   \frac{d\,\tilde c_{BB}}{d\ln\mu}
   &=  \frac{17y_t^2}{96\pi^2}\,c_{tt} 
    + \frac{11\alpha_s^2}{2\pi^2}\,\tilde c_{GG}
    + \frac{9\alpha_2^2}{8\pi^2}\,\tilde c_{WW}
    + \frac{95\alpha_1^2}{24\pi^2}\,\tilde c_{BB} \,,
\end{align}
where we defined
\begin{align}\label{eq:Vcouplings}
   \tilde c_{GG} 
   &= c_{GG} + \frac12\,\text{Tr} \left( \bm{c}_u + \bm{c}_d - 2\,\bm{c}_Q \right) , \\
   \tilde c_{WW} 
   &= c_{WW} - \frac12\,\text{Tr} \left( 3\,\bm{c}_Q + \bm{c}_L \right) , \\
   \tilde c_{BB} 
   &= c_{BB} + \text{Tr} \left( \frac43\,\bm{c}_u + \frac13\,\bm{c}_d - \frac16\,\bm{c}_Q 
    + \bm{c}_e - \frac12\,\bm{c}_L \right) ,
\end{align}
The running of the physical ALP-fermion couplings
\begin{align}\label{eq:cqq}
c_{qq}=c_q-c_Q, \qquad c_{ee}=c_e-c_L\,,
\end{align}
with $q=t,u,d$ , is given by
\begin{align}\label{cttRGE}
   \frac{d\, c_{tt}}{d\ln\mu}&=\frac{9 y_t^2}{16\pi^2}\,c_{tt}
    + \frac{2\alpha_s^2}{\pi^2}\,\tilde c_{GG} 
    + \frac{9\alpha_2^2}{16\pi^2}\,\tilde c_{WW}
    + \frac{17\alpha_1^2}{48\pi^2}\,\tilde c_{BB} \,\\
  \frac{d\, c_{uu}}{d\ln\mu}&=\frac{3 y_t^2}{8\pi^2}\,c_{tt}
    + \frac{2\alpha_s^2}{\pi^2}\,\tilde c_{GG} 
    + \frac{9\alpha_2^2}{16\pi^2}\,\tilde c_{WW}
    + \frac{17\alpha_1^2}{48\pi^2}\,\tilde c_{BB} \,.\\
  \frac{d\, c_{dd}}{d\ln\mu}&=\frac{3 y_t^2}{8\pi^2}\,c_{tt}
    + \frac{2\alpha_s^2}{\pi^2}\,\tilde c_{GG} 
    + \frac{9\alpha_2^2}{16\pi^2}\,\tilde c_{WW}
    + \frac{5\alpha_1^2}{48\pi^2}\,\tilde c_{BB} \,.\\
  \frac{d\, c_{\ell\ell}}{d\ln\mu}&=\frac{3 y_t^2}{8\pi^2}\,c_{tt}
    + \frac{9\alpha_2^2}{16\pi^2}\,\tilde c_{WW}
    + \frac{19\alpha_1^2}{16\pi^2}\,\tilde c_{BB} \,.
\end{align}
Here, the running of the ALP-charm coupling $c_{cc}$ and the ALP-strange coupling $c_{ss}$ is described by the beta function of $c_{uu}$ and $c_{dd}$, respectively. 

Matching contributions for the ALP couplings to gauge bosons at the electroweak scale can be written as
\begin{equation}\label{CVVmatch}
\begin{aligned}
   \tilde c_{GG}(\Lambda) 
   &= c_{GG} + \frac12 \sum_q c_{qq}(\Lambda) \,, \\[-2mm]
   \tilde c_{WW}(\Lambda) 
   &= c_{WW} - \frac12\,\text{Tr}\,\Big[ 3\bm{c}_U(\Lambda) + \bm{c}_E(\Lambda) \Big] \,, \\[1mm]
   \tilde c_{BB}(\Lambda) 
   &= c_{BB} + \sum_f\, N_c^f\, Q_f^2\,c_{ff}(\Lambda) 
    + \frac12\,\text{Tr}\,\Big[ 3\bm{c}_U(\Lambda) + \bm{c}_E(\Lambda) \Big] \,.
\end{aligned}
\end{equation}
Matching contributions to the physical ALP-fermion couplings at the electroweak \,scale can be written for $f=u,d$ read
\begin{align}\label{cffmatch}
   \Delta\bm{c}_{ff}(\mu_w) 
   &= -\frac{3 y_t^2}{8\pi^2}\,c_{tt}\,T_3^f\,
    \ln\frac{\mu_w^2}{m_t^2}\,\mathbbm{1} \notag\\
   &\quad\mbox{}-\frac{3\alpha^2}{16\pi^2 s_w^2}\bigg[\frac{c_{WW}}{s_w^2} \left(\ln\frac{\mu_w^2}{m_W^2} + \frac12\right) + \frac{4c_{\gamma Z}}{c_w^2}Q_f(T_3^f-2Q_fs_w^2) \left( \ln\frac{\mu_w^2}{m_Z^2} + \frac32 \right) \notag\\
  &\qquad\quad+ \frac{4c_{ZZ}}{c_w^4 s_w^2} \left(s_w^2Q_f^2
   +(T_3^f-Q_fs_w^2)^2\right) \left(\ln\frac{\mu_w^2}{m_Z^2} + \frac12 \right)\bigg]-\delta c_{DD}(\mu_w)\mathbbm{1}
\end{align}
where the last line includes a contribution from $W-$loops with internal top quarks that is only present for the matching of ALP couplings to down-type at the weak scale, where
\begin{align}\label{eq:KDeff}
  \delta c_{DD}(\mu_w)
   = \frac{y_t^2}{16\pi^2}\,|V_{tD}|^2 \bigg(&\frac{c_{tt}(\mu)}{4}\Big(-1+2\ln \frac{\mu_w^2}{m_t^2}-6\frac{1-x_t+\ln x_t}{(1-x_t)^2}  \Big) \notag\\
   &-\frac{3\alpha }{2\pi s_w^2}c_{WW}\frac{1-x_t+x_t\ln x_t}{(1-x_t)^2}\bigg)
\end{align}
and $D=d,s,b$.

Running below the electroweak scale ALP couplings to gauge couplings only run at 2-loop order and can be neglected. The running of ALP-fermion couplings below the electroweak scale is given by
\begin{align}\label{EWtoQCDrunning}
 \frac{d\, c_{qq}}{d\ln\mu}&=\frac{\alpha_s^2}{\pi^2}\tilde c_{GG}+\frac{3\alpha^2}{4\pi^2}Q_q^2\tilde c_{\gamma\gamma}\,,\\
  \frac{d\, c_{ee}}{d\ln\mu}&=\frac{3\alpha^2}{4\pi^2}\tilde c_{\gamma\gamma}\,.
\end{align}
There are threshold corrections to the ALP couplings to gauge bosons whenever a new quark or lepton is integrated out so that
\begin{align}\label{CVVbelowQCD}
\tilde c_{GG}&=c_{GG} +\frac12 \sum_q c_{qq}(\mu)\theta(\mu-m_q)\\
\tilde c_{\gamma\gamma}&=c_{\gamma\gamma} + \sum_f N_c^f c_{ff}(\mu)\theta(\mu-m_f)
\end{align}
And evaluating the ALP-fermion couplings at the QCD scale requires a step-wise matching and running across flavor thresholds. 

The coefficients entering \eqref{LlowE} are now obtained by first running the ALP couplings from the UV scale to the electroweak scale by solving the system of RGEs at the weak scale~\eqref{RGEs2l} and~\eqref{cttRGE} (as well as the running of the SM gauge couplings and the top Yukawa coupling) and adding the matching contributions~\eqref{CVVmatch} and \eqref{cffmatch}. The ALP couplings at the QCD scale are then determined by step-wise running below the electroweak scale via~\eqref{EWtoQCDrunning}, taking into account the scale dependence of the ALP-gauge boson couplings ~\eqref{CVVbelowQCD} and of $\alpha$ and $\alpha_s$.\\
The effective ALP coupling to photons below the QCD scale can be written as 
\begin{equation}\label{eq:CgagaQCD}
\begin{aligned}
   c_{\gamma\gamma}^\text{eff}(m_a< \mu_0)
   &= c_{\gamma\gamma}(\mu_0) - \left(\frac53+\frac{m_\pi^2}{m_\pi^2-m_a^2}\frac{m_d-m_u}{m_u+m_d}\right)\,c_{GG}(\mu_0) \\
   &\qquad - \frac{m_a^2}{m_\pi^2-m_a^2} 
     \frac{c_{uu}(\mu_0)-c_{dd}(\mu_0)}{2}+ \sum_{q=c,b} 3\, Q_q^2\,c_{qq}(\mu_0)\,B_1(m_q)\\
    &\qquad  + \sum_{\ell=e,\mu,\tau} c_{\ell\ell}(\mu_0)\,B_1(m_\ell) \\
 & = \, c_{\gamma\gamma}(\Lambda) - (1.92\pm 0.04)\,c_{GG}(\Lambda)+\mathcal{O}\bigg(\tau_a, \frac{m_a^2}{m_f^2}\bigg)\end{aligned}
\end{equation}
because the loop function $B_1(\tau_f)\approx -m_a^2/(12m_f^2)$ for $m_f\gg m_a$. The coefficient of $c_{GG}$ in the second line includes the NLO corrections calculated in \cite{GrillidiCortona:2015jxo}.

\section{Axion models}
The couplings in~\eqref{eq:Vcouplings} and ~\eqref{eq:cqq} are general Wilson coefficients for a pseudoscalar with approximate shift symmetry coupled to SM particles at the UV scale. Various explicit axion models have been proposed, e.g. to explain the small CP phase in QCD, and in the following we give the relation between axion couplings in these models and these Wilson coefficients.

Arguably the simplest implementation of the QCD axion is the KSVZ model with a single coupling between the axion and gluons in the UV. The complete theory needs additional color charged particles that generate this coupling, but in terms of the EFT above the electroweak scale one only has
\begin{align}
c_{GG}\big\vert_\text{KSVZ}=\frac12\,.
\end{align}
Alternatively one can generate the coupling of the axion to gluons via quark loops as in the case of the DSFZ model. In this case, one has in the UV
\begin{align}
c_{GG}\big\vert_\text{DSFZ}&=\frac12\,,\notag\\
c_{\gamma\gamma}\big\vert_\text{DSFZ}&=\frac{3}{16}\,,\notag\\
c_{uu}\big\vert_\text{DSFZ}&=\frac13\cos^2\beta\,,\notag\\
c_{dd}\big\vert_\text{DSFZ}&=c_{ee}\big\vert_\text{DSFZ}=\frac13\sin^2\beta\,.
\end{align}
Alternatively, the DSFZ model can be realised with the couplings
\begin{align}
c_{GG}\big\vert_\text{DSFZII}&=\frac12\,,\notag\\
c_{\gamma\gamma}\big\vert_\text{DSFZII}&=\frac{1}{8}\,,\notag\\
c_{uu}\big\vert_\text{DSFZII}&=c_{ee}\big\vert_\text{DSFZII}=\frac13\cos^2\beta\,,\notag\\
c_{dd}\big\vert_\text{DSFZII}&=\frac13\sin^2\beta\,.
\end{align}

\section{Big Bang Nucleosynthesis and astrophysical constraints}

For completeness, we provide the equations relevant for the computation of constraints from Big Bang Nucleosynthesis (BBN) in terms of the low energy theory presented in Section~\ref{sec:theory} here. The abundance of elements during BBN is sensitive to variations in the neutron-proton mass difference which changes the ratio of neutrons to protons in the early Universe. If there is a non-zero $\theta$ angle the neutron-proton mass difference becomes $\theta$ dependent and the binding energies of nucleons are increased, resulting in constraints $\theta\lesssim 0.1$ during BBN~\cite{Blum:2014vsa, Lee:2020tmi}. Even in the absence of a $\theta$ angle variations of the nucleon masses from quadratic interactions can change the neutron-proton mass difference~\cite{Coc:2006sx, Flambaum:2007mj, Stadnik:2015kia, Lee:2020tmi, Bouley:2022eer}. The strongest constraint on ALP interactions arises from the change in the Helium abundance, which can be written as~\cite{Bouley:2022eer}
\begin{align}
\frac{\Delta Y_p}{Y_p}=&\frac{\Delta X_{n}(a_W)}{X_n(a_W)}+\frac{\Gamma_n}{H(a_\text{\tiny{BBN}})}\frac{\Delta B_D}{B_D} \\
&-\int_{a_\text{W}}^{a_\text{\tiny{BBN}}}\frac{da'}{a'}\frac{\Gamma_n}{H(a')}\left(\frac{6g_A}{1+3g_A}\Delta g_A +\frac{2\Delta G_F}{G_F}+5\delta_e +x\frac{P'(x)}{P(x)}\Big(\delta_{\Delta M}-\delta_e\Big) \right)\,,\notag
\end{align}
where $X_n(a_W)$ is the neutron abundance at weak freeze-out, $B_D$ the deuterium binding energy, $H(a')$ the Hubble constant in dependence of the scale factor $a'$, 
 $g_A$ is the weak axial coupling, the neutron decay width reads 
 \begin{align}
 \Gamma_n=\frac{1+3g_A^2}{2\pi^3}G_F^2m_e^5 P(x),
 \end{align}
and  
\begin{align}
P(x)=\frac{1}{60}\left((2x^4-9x^2-8)\sqrt{x^2-1}+15x \ln (x+\sqrt{x^2-1})\right)
\end{align}
and $P'(x)$ are the phase space factor and its derivative
evaluated at $x=\Delta M_N/m_e$. The neutron abundance at weak freeze-out can be written in the instantaneous approximation (neglecting variations of the freeze-out temperature $T_W$)
\begin{align}
\frac{\Delta X_{n}(a_W)}{X_n(a_W)}=-\frac{\Delta M_N}{T_W}\delta_{\Delta M}\,\approx -1.358\, \delta_\pi(a)
\end{align}
The dependence of the deuterium binding energy on the pion mass has been obtained in~\cite{Mosquera:2010mv}, so that
\begin{align}
\frac{\Delta B_D}{B_D} = -1.825\,\delta_\pi(a)\,.
\end{align}
In order to express the variation of the weak axial nucleon coupling on the ALP field can be obtained from NNLO heavy baryon ChPT~\cite{Bouley:2022eer}
\begin{align}
\frac{\Delta g_A}{g_A}= -0.008 \,\delta_\pi(a)\,,
\end{align}
and the variation of the Fermi constant is 
\begin{align}
\frac{\Delta G_F}{G_F}=\delta_\alpha(a)\,.
\end{align}
The result is directly sensitive to the square of the field values $a^2/f^2$, which depend on the cosmic history of the ALP field and a recent analysis for quadratically interacting fields can be found in ~\cite{Bouley:2022eer}, where one can also find constraints from Lyman-$\alpha$ measurements~\cite{Rogers:2020ltq}, ultrafaint dwarf galaxies~\cite{Dalal:2022rmp} and black hole superradiance~\cite{Baryakhtar:2020gao}.

\bibliography{biblio}
\bibliographystyle{jhep}

\end{document}